\documentclass{aa}
\usepackage{txfonts}
\usepackage{graphics}
\usepackage{amssymb}
\newcommand{\kms}{km\,s$^{-1}$}
\newcommand{\Lfir}{$L_{\rm fir}$}
\newcommand{\Msol}{M$_{\odot}$}
\newcommand{\Lsol}{L$_{\odot}$}

\newcommand{\Tex}{$T_{\rm ex}$}

\newcommand{\degs}{$^{\circ}$}
\newcommand{\pad}{.\hskip-2pt$^\circ$}
\newcommand{\pam}{.\hskip-2pt$^{\prime}$}
\newcommand{\pas}{.\hskip-2pt$^{\prime\prime}$}
\newcommand{\pats}{.\hskip-2pt$^s$}
\newcommand{\magn}{.\hskip-2pt$^m$}
\newcommand{\gsim}{\;\lower.6ex\hbox{$\sim$}\kern-7.75pt\raise.65ex\hbox{$>$}\;}
\newcommand{\lsim}{\;\lower.6ex\hbox{$\sim$}\kern-7.75pt\raise.65ex\hbox{$<$}\;}
\begin{document}

\title{A star cluster at the edge of the Galaxy 
\footnote {Partly based on observations collected at
the European Southern Observatory, Chile.}}

\author{J. Brand
         \inst{1}
         \and
	 J.G.A. Wouterloot
	 \inst{2}
	  }

\offprints{J. Brand (brand@ira.inaf.it)}

\institute{
INAF - Istituto di Radioastronomia, Via P. Gobetti 101, I-40129 Bologna, Italy 
 \and
 Joint Astronomy Centre, 660 N. A'ohoku Place, University Park, 
 Hilo, HI 96720, USA
}

\date{Received ;accepted }

\abstract
{This paper is part of our ongoing study of star formation in the (far-) outer 
Galaxy.}
{Our goal in this paper is to study stars and molecular gas in the direction 
of IRAS06145+1455 (WB89-789). 
The kinematic distance of the associated molecular cloud is 11.9~kpc. With a 
galactocentric distance of $\sim 20.2$~kpc, this object is at the edge of the
(molecular) disk of the Galaxy.}
{We use near-IR (J, H, K), molecular line-, and dust continuum observations.
}
{The near-IR data show the presence of an (embedded) cluster of about
60 stars, 
with a radius $\sim 1.3$~pc and an average stellar surface density 
$\sim 12$~pc$^{-2}$. We find at least 14 stars with NIR-excess, 3 of which are
possibly Class~I objects. The cluster is embedded in a $\sim 1000$~\Msol\ 
molecular/dust core, from which a molecular outflow originates. 
The temperature of most of the outflowing gas is $\lsim 40$~K, and 
the total mass of the swept-up material is $\lsim 10$~\Msol. Near the center of
the flow, indications of much higher temperatures are found, probably due to 
shocks. 
A spectrum taken of one of the probable cluster members shows a tentative 
likeness to that of a K3~III-star (with an age of at least 20~Myr). If
correct, this would confirm the kinematic distance.
} 
{This cluster is the furthest one from the Galactic center yet detected.
The combination of old and recent activity implies that star formation 
has been going on for at least 20~Myr, which is difficult to understand 
considering the location of this object, 
where external triggers are either
absent or weak, compared to the inner Galaxy. This suggests that once 
star formation is occurring, later generations of stars may form through 
the effect of the first generation of stars on the (remnants of) the 
original molecular cloud.}

\authorrunning{Brand \& Wouterloot}
\titlerunning{Star cluster at the edge of the Galaxy}
\maketitle

\section{Introduction \label{intro}}
The H{\sc i} in our Galaxy extends out to galactocentric distances $R$ of at 
least 24-25~kpc (e.g., Wouterloot et al.~\cite{wbbk}; McClure-Griffiths et 
al.~\cite{mcclure}). From high-sensitivity 21-cm observations, Knapp et al. 
(\cite{knapp}) found H{\sc i} out to $R \approx 50$~kpc (for a flat rotation
curve and R$_0$=8.5~kpc).
Molecular material 
and associated sites of star formation do not appear at such large distances, 
however. Star formation in the 
Galaxy has been observed to occur out to  
$R\approx 20$~kpc (e.g., Fich \& Blitz~\cite{fich84}; 
Wouterloot et al.~\cite{wbh}; Kobayashi \& Tokunaga~\cite{koba}; Santos et 
al.~\cite{santos}; Snell et al.~\cite{snell}). Wouterloot \& Brand 
(\cite{wb89}) detected many molecular clouds (hence star formation sites) 
towards IRAS sources in the outer regions of the disk.
Thus both star formation and star formation reservoirs (i.e., molecular 
clouds) are present 
at large distances from the Galactic center, but not as far out as the 
atomic hydrogen. This suggests that beyond about 
$R\approx 20$~kpc, conditions are not favorable to transforming H{\sc i} into 
H$_2$ and then into stars. The cause of this is not known (in general the 
formation of H$_2$ clouds from H{\sc i} is not yet a settled manner, 
but it must have to do with the physical environment presented by the
interstellar medium).

In the far-outer Galaxy ($R>16$~kpc; hereafter FOG) the physical environment
differs from that in the inner Galaxy (for a summary see Brand \& 
Wouterloot~\cite{bw95}), 
the effects of which may influence the formation of molecular clouds and the 
star formation process within them. Also, both the volume density of the
molecular (and atomic) gas and the strength of spiral density waves 
are much reduced in the outer Galaxy. 
Even if star formation is {\it independent} of cloud formation and is 
to be initiated or enhanced by cloud-cloud collisions or triggered by
supernovae or density waves, then star formation activity 
in the outer Galaxy would be expected to be significantly lower than 
in the inner Galaxy.
To study the process of star formation and its end products in the outer parts 
of the galactic molecular disk, 
we observed a selection of FOG clouds in the NIR to directly
detect the embedded stellar population. FOG clouds are at large enough $R$
for the physical conditions of the ISM to be different, yet those in the 
2$^{nd}$ and 3$^{rd}$ quadrants are close enough
for the star-forming cores to be resolved. 

Here we  present 
the most distant object studied with these observations: 
\object{WB89-789} (\object{IRAS06145+1455}; $\alpha_{2000}=06^h17^m$24\pats2, 
$\delta_{2000}$=+14\degs 54\arcmin 42\arcsec) 
at ({\sl l,b} = 195\pad82, $-$0\pad57),
which is located at $R \approx 20.2$~kpc, $d \approx 11.9$~kpc (Brand \& 
Wouterloot~\cite{bw94}; hereafter BW94). The luminosity of the
IRAS source (\Lfir$\approx 1.9 \times 10^4$~\Lsol) is consistent with 
a single star of
type B0.5V; it is embedded in a small (equivalent radius $r_e \approx 5.7$~pc, 
and
$M \approx 5.6 \times 10^3$~\Msol\ [BW94]) cloud. An H$_2$O maser was detected
towards this object (Wouterloot et al.~\cite{wbf}), indicating that this is
an active star-forming region. No radio continuum emission was found 
within 4$^{\prime}$ of the location of the IRAS source in 
our VLA A-array observations at 3.6~cm (Nov. 1993; {\it unpublished}). 
With a 4$\sigma$ detection limit of 
$\sim$ 3~mJy we would have been able to detect an H{\sc ii} region ionized 
by a B0.5V star or earlier.

\section{Observations and data reduction \label{obsred}}

\subsection{Near-infrared}
The data presented here were obtained on February 15, 1995, with the
ESO 2.2-m telescope at La Silla (Chile). Images in J, H, and K-bands were
taken with the IRAC-2 camera and objective C, which resulted in a scale of
0\pas49/pixel. The detector was a $256 \times 256$~pixel$^2$ NICMOS-III. The 
total instantaneous field-of-view is therefore $\sim 2.1 \times 
2.1$~arcmin$^2$. During the observations the seeing was 0\pas9.

From a quick K-band exposure ($20 \times 3$~seconds) centered on the WB 89-789 
IRAS point source position, a group of stars was seen to be small enough in 
extent to
allow an observing procedure where it was placed at the center of
each quadrant of the CCD, as well as near the center of the CCD itself.
Integrations in the J-, H-, and K-bands were then made for
$60 \times 2$~seconds, resulting in 5 separate images per band. 
After correction with 
a bad-pixel mask, each image was reduced
separately by subtracting a starless sky image derived by median averaging
from the other 4 exposures, and dividing by the flat field (see below).
The reduced images were then averaged to get the final frame, which has an
effective integration time of 10~minutes. 
Immediately after these ``ON-source'' observations a comparison (OFF-) field
was observed, at the same galactic latitude, but shifted towards smaller
longitude by 0\pad2 ($\approx 42$~pc at the distance of WB89-789).
To construct flat fields, dome exposures in all three bands were taken at the
start and at the end of each night.

We observed 5 standard stars during the night, at different airmass. Each 
standard star was observed 
at 5 different positions on the CCD (with the same observing pattern as used 
for our object) in all 3 bands. Each observation was reduced individually and
the derived photometric zero points (ZPs) were averaged, after which the ZP
at the airmass of the WB89-789 observation was obtained through linear 
interpolation.

The standard star observations revealed a non-uniform CCD, with some quadrants
``hotter'' (i.e., registering more electrons for the same number of infalling 
photons) than others. 
This non-uniformity 
will increase the random error in the photometry of the science objects. 
In part this is accounted for by the average ZP, as the standards were 
observed in the same way as the 
science object, hence the average ZP partly 
takes into account the differences between the quadrants.

The mosaiced images in J, H, and K were aligned and truncated 
to show the same region of the sky. 
Stars were identified in each frame with the IRAF task DAOFIND, and
photometry was performed through DAOPHOT, using a point-spread function
determined from 5 relatively bright and isolated stars. Fitted stars
were subtracted from the images, after which DAOFIND was run again on the
residual frame. This procedure was repeated until no more stars were found
(usually after 2 or 3 iterations). Derived magnitudes were then corrected to
take into account the difference between the
point spread fitting radius (3 pixels) and the aperture radius used (30 pixels)
on the standard stars to derive the ZP; this correction amounts to $-$0\magn34,
$-$0\magn27, and $-$0\magn24 for J, H, and K, respectively. 

\subsection{Millimeter-lines}
{\bf JCMT}\hfill\break\noindent
On January 24, 2003, we used the 15-m James Clerk Maxwell Telescope (JCMT) on
Mauna Kea (Hawaii) to
map the emission of $^{12}$CO(2--1) towards WB89-789 in the raster 
(on-the-fly) observing mode. The JCMT beam size at 230~GHz
is 22\arcsec; observations were made on a 10\arcsec\ raster, using an 
autocorrelator spectrometer at a velocity resolution of 0.2~\kms. 
An 
off-position was used at offset 600\arcsec\ in right ascension from the IRAS 
source position.
The typical rms noise level in the spectra is 0.2~K ($T^*_{\rm A}$). 

\smallskip
On August 7 and 8, 2003, the JCMT was used to simultaneously observe 
$^{13}$CO(2--1) and C$^{18}$O(2--1) emission in a 80\arcsec $\times$
80\arcsec\ region on a 10\arcsec-raster towards WB89-789 using 
frequency-switching over 8.2~MHz.
The typical rms noise level in these spectra is 0.13~K ($T^*_{\rm A}$).
A comparison with separate (i.e., non-simultaneous) $^{13}$CO(2--1) 
observations of line calibrators shows that our $^{13}$CO(2--1) intensities 
must be corrected by multiplication with a factor of 0.88.

\smallskip
On January 5, 2006, we used the JCMT to observe the $^{12}$CO(3--2) transition
towards WB89-789, in essentially the same region covered by the $^{12}$CO(2--1)
observations. Observations were made on a grid with 6\arcsec\ spacing; the
beam size at 345.796~GHz is 14\arcsec. The velocity resolution is 0.136~\kms,
and typical rms noise level is 0.6~K ($T^*_{\rm A}$). Pointing was found to 
be accurate within 1\pas8.
The line-data were reduced and analyzed with the programs CLASS and GRAPHIC, 
which are part of the GAG-software package developed by the Obs. de Grenoble 
and IRAM-Grenoble.

\smallskip\noindent
{\bf IRAM 30-m}\hfill\break\noindent
WB89-789 was observed in CS J=2--1, 3--2, and 5--4 with the 
IRAM 30-m telescope at Pico Veleta (Granada, Spain) in the period 
July 24 -- 26, 1991. Maps were made of 
thirteen positions, with 15\arcsec\ spacing.
The region mapped was sufficient to
cover all CS emission.  The telescope beamwidth at these frequencies (98,
147, and 245~GHz) is respectively 26\arcsec, 16\arcsec, and 11\arcsec. 
All intensities are on a $T^*_{\rm A}$ scale.
For CS(2--1) and (3--2) we
used a 2$\times$128-channel filterbank of 100~kHz per channel, resulting in a
resolution of 0.31 and 0.20~kms$^{-1}$, respectively; CS(5--4) was observed 
with a resolution of 584~kHz (0.71~kms$^{-1}$). Observations were made by 
position switching against an off-position 30\arcmin\ W;
the typical rms noise level in the spectra at the central position is 0.05~K. 
All spectra were reduced with the Grenoble CLASS software. 

\subsection{Millimeter-continuum}
{\bf JCMT}\hfill\break\noindent
On April 10, 2002, we simultaneously observed the 450~$\mu$m and 850~$\mu$m
emission towards WB89-789 with the Submillimetre Common-User Bolometer Array 
(SCUBA; Holland et al.~\cite{holland}) at the JCMT. A standard 64-point jiggle
map was made, which covers an area 2\pam3 in diameter; hence the observations 
are limited to the region immediately around the IRAS source. 
Calibration was performed by using fits to the time-dependence of skydips and
the nearby Caltech Sub-millimeter Observatory (CSO) 
225~GHz optical depth data, and average ratios of $\tau$(450~$\mu$m) 
and $\tau$(850~$\mu$m) to $\tau$(225~GHz). 
The final transition from instrumental 
parameters to Jy/beam was derived from similar maps of Mars. 
The half-power beam size at 450~$\mu$m and 850~$\mu$m are 
9\arcsec\ and 15\arcsec, respectively. 
The rms values 
in the WB89-789 maps are 450~mJy/beam (at 450~$\mu$m) and 20~mJy/beam 
(at 850~$\mu$m), respectively.
The data were reduced with the program SURF 
(Jenness \& Lightfoot~\cite{jenness}).
We found that away from the continuum source, the background emission level
at both wavelengths was not around zero, but somewhat higher. We have 
corrected this by subtracting 1~Jy and 0.05~Jy from all pixel values in the 
450~$\mu$m and 850~$\mu$m maps, respectively.

\smallskip\noindent
{\bf SEST}\hfill\break\noindent
Between August 17 and 22, 2003, the region around WB89-789 was observed in the
1.2-mm continuum with the 37-channel bolometer array SIMBA (SEST IMaging 
Bolometer Array) at the 15-m Swedish-ESO Submillimetre Telescope (SEST) at 
ESO-La Silla, Chile. 
A region of size 600\arcsec\ $\times$ 600\arcsec\ (azimuth $\times$ 
elevation) was scanned at a rate of 80\arcsec/s; the total integration time 
per map was about 12~minutes. Atmospheric opacity was determined from 
skydips, which were taken every 2~hours; values at the zenith ranged between
0.15 and 0.45. The data were transformed from counts/beam to mJy/beam by 
making similar maps of Uranus once per day. The average conversion factor 
during this observing period is 58$\pm$6~mJy/count. 
\hfill\break\noindent
In total, 8 maps were obtained. Each map was reduced and 
calibrated individually, after which they were averaged. The rms in the
final image is 15~mJy/beam.
Pointing of the SEST was checked by observing a strong continuum source every 
2~hours. The half-power beam width of the SEST at 1.2~mm is 23\arcsec.
Data were reduced with the program MOPSI, written by Robert Zylka (IRAM, 
Grenoble), following the instructions from the SIMBA Observers Handbook 
(Version 1.9, 9 Feb. 2003).

\subsection{Spectroscopy}
On December 7, 2004, we used the DOLORES spectrograph at the Telescopio 
Nazionale Galileo (TNG) at La Palma (in service observing-mode) to take a 
spectrum of one of the stars of the 
cluster. Six separate integrations of 2220~sec each were taken using the
high-resolution spectrograph and HR-B Grism nr.~5 (dispersion 0.875\AA\ per 
pixel), which resulted in a 
wavelength coverage of about 3200\AA\--4900\AA, and (with a 1\arcsec\ slit) 
in a spectral resolution of $\sim$3.2\AA. 
Bias frames were taken on the same night, and flat fields and He and Ar
calibration-lamp spectra were taken the night before and after. As a flux
calibrator, Hiltner600 was observed on the same night.
\hfill\break\noindent
Data reduction was performed in IRAF, using standard packages. Bias frames
were subtracted from all science and flat field frames. An average 
normalized flat field was constructed, which was corrected for bad pixels and
then divided into the star frames. Wavelength calibration was performed with
the Ar-lamp spectra, which in the wavelength range in question only has 8
useful (i.e., non-blended) lines, none below 3780\AA, and is therefore not
very accurate. 

\begin{figure*}[tp]
\resizebox{8cm}{!}{\includegraphics{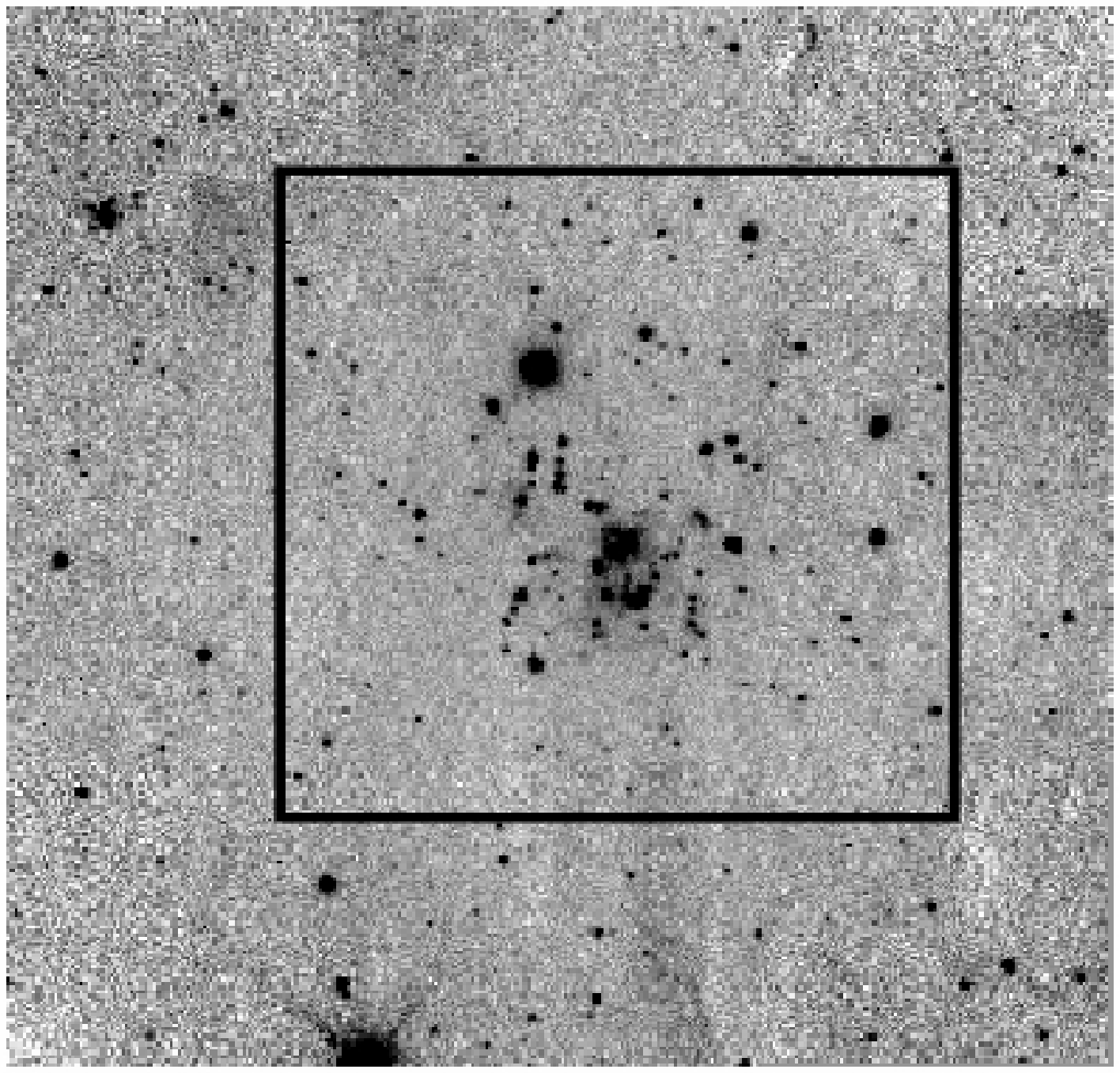}}
\resizebox{6cm}{!}{\includegraphics{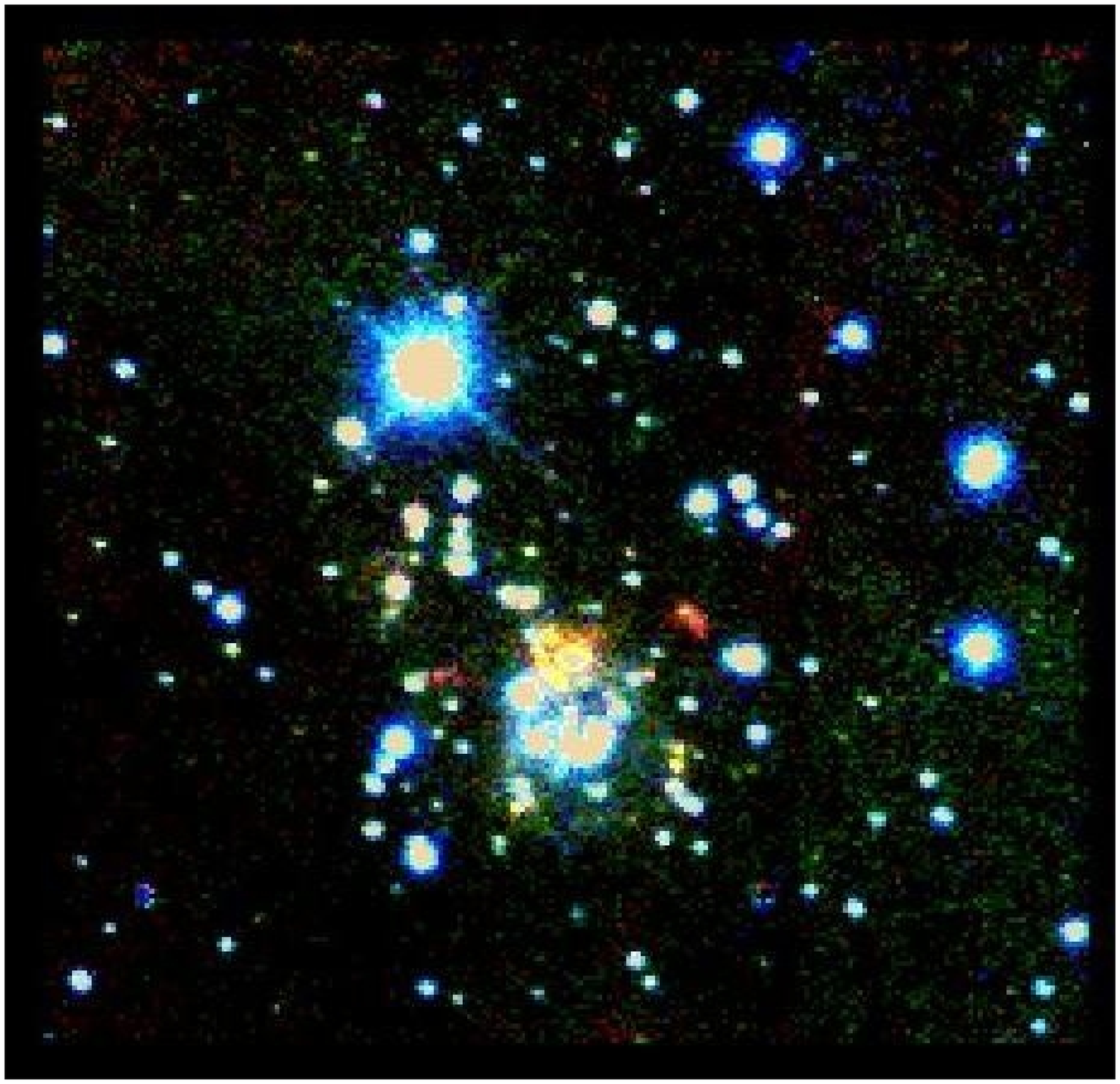}}
\caption[]{{\bf a}\ (left). 
K-band image of the region around WB89-789 (IRAS06145+1455). 
The area of sky visible here is $\sim 3 \times 3$~arcmin$^2$. The region 
outlined by the black box has a size of $\sim 1.7 \times 1.7$~arcmin$^2$, and 
is shown on the right. North is up, East is left.
{\bf b}\ (right). False-color image (J=blue, H=green, K=red) of a
$\sim 1.7 \times 1.7$~arcmin$^2$ region around WB89-789 
}
\label{koverview}
\end{figure*}

\section{Results}

\subsection{Stars}

\subsubsection{Photometry, clustering, and pre-main~sequence stars}

In Fig.~\ref{koverview}a we show the K-band mosaic, while 
a false-color (JHK) image of the area around WB89-789 (the region inside the 
box in Fig.~\ref{koverview}a) is shown in Fig.~\ref{koverview}b.
Just below the center of the image we see a group of stars embedded in 
nebulosity; the 
IRAS source is located in the northern half of the nebulosity. 
In this 1\pam7 $\times$ 1\pam7 area DAOFIND detects 173, 179, and 161 stars in 
the J-, H-, and K-bands, respectively. 
The distributions of these stars in bins of 1 magnitude peak at 
$m_{\rm J} \approx 20$
(faintest star found 19.96$\pm$0.24), $m_{\rm H} \approx 18$ (19.59$\pm$0.54), 
and $m_{\rm K} \approx 17$ (18.52$\pm$0.50), respectively. 
Signal-to-noise ratios are reached  
of 5--10, i.e., photometric accuracy of 0\magn1--0\magn2, at $m_{\rm K} = 
16.5 - 17.5$, which corresponds to main sequence [ms] spectral types A0--5 
(2--3~\Msol)
for a distance of 10~kpc and visual extinction $A_{\rm V}$=10~mags.

\begin{figure}[tp]
\resizebox{7cm}{!}{\includegraphics{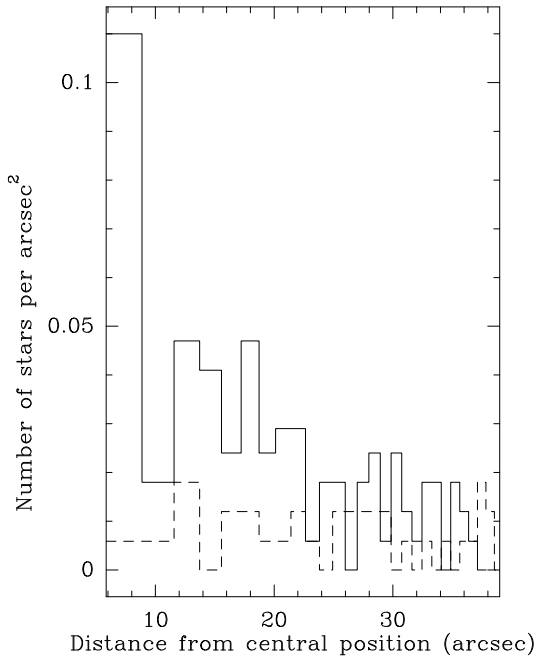}}
\caption[]{Comparison between the star densities in the WB89-789 field (drawn)
and in a field centered at a nearby off-position (dashed). 
Stars were counted on the K-images in concentric rings of equal area.
In the WB89-789 field the rings were centered on the nebulosity
(Fig.~\ref{koverview}b). An excess of stars 
in the WB89-789 field is clearly visible.}
\label{onoffcomp}
\end{figure}

\begin{figure*}[tp]
\resizebox{6cm}{!}{\rotatebox{270}
{\includegraphics{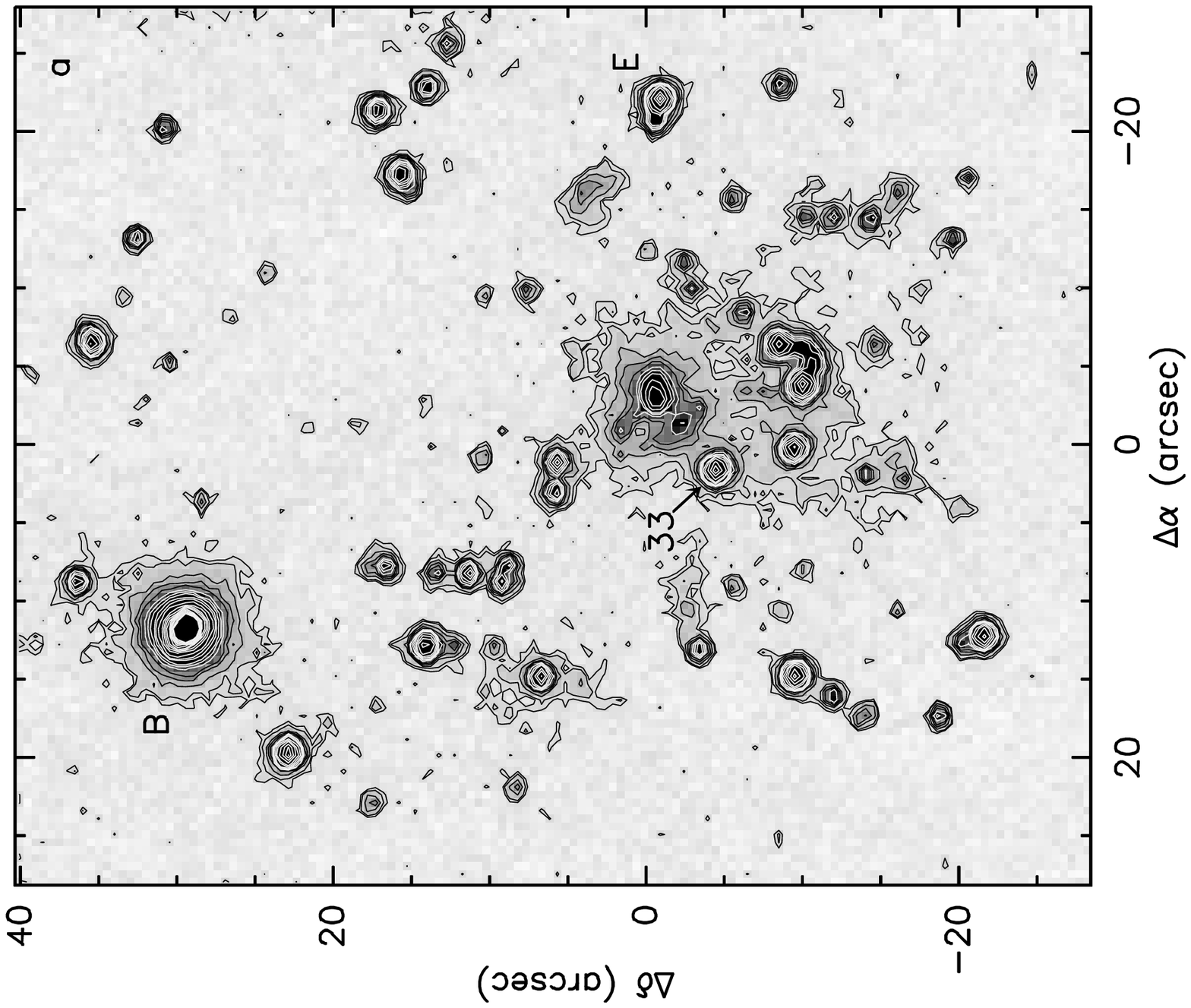}}}
\hspace{1cm}
\resizebox{7.5cm}{!}{\rotatebox{270}
{\includegraphics{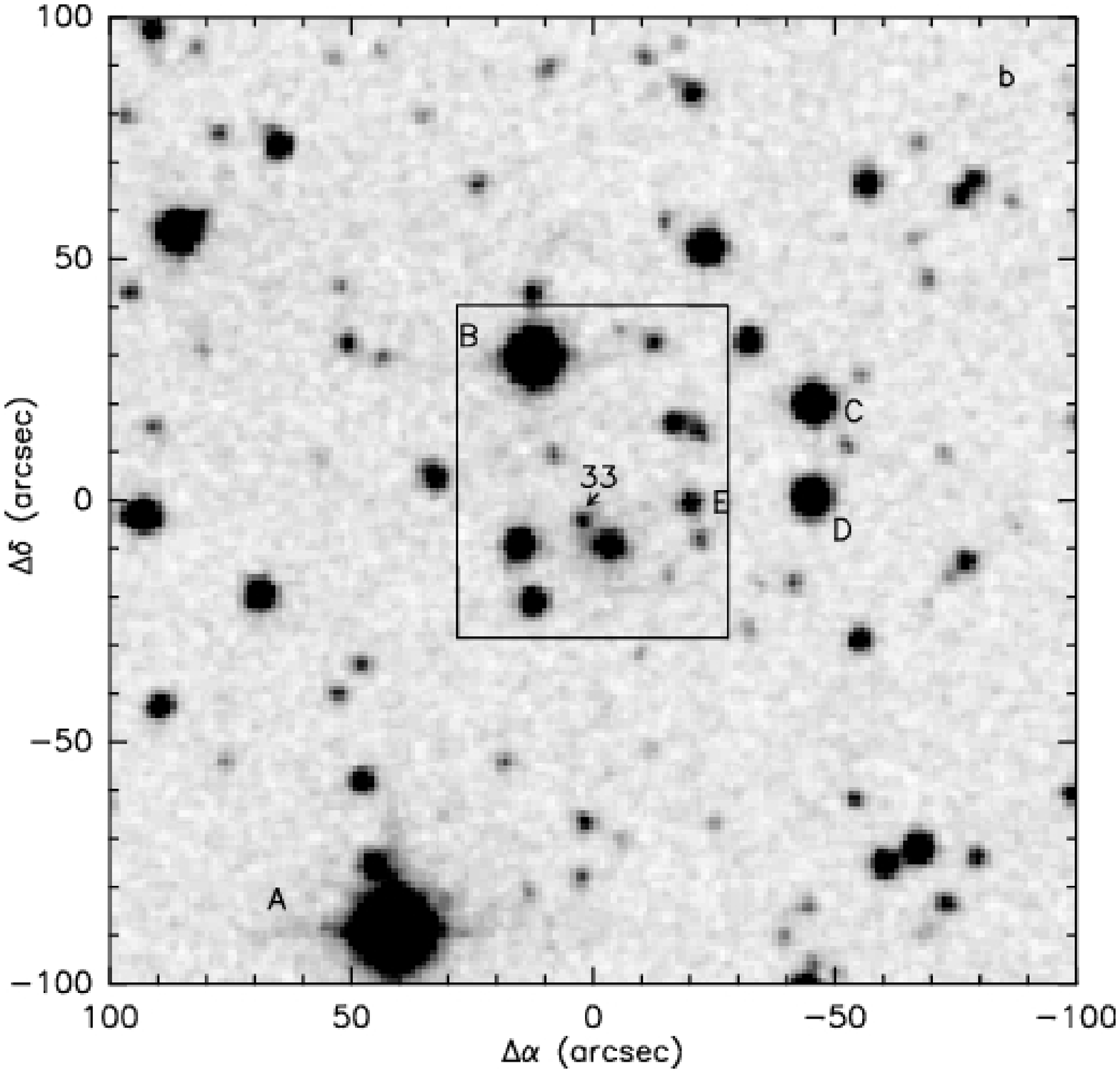}}}
\caption[]{
{\bf a.}\ 
Contour plot (K-frame) of that part of the final mosaic where all 5 
ON-source images overlap. Sixty-eight stars in this area have been detected 
in all three bands (J, H, and K).
Several stars are labeled, to identify the field in panel~b. Among
them is star 33, of which a spectrum has been taken (Sect.~2.4). 
North is up, East is left. The (0,0)-offset indicates the nominal position of 
the IRAS source.
\hfill\break\noindent
{\bf b.}\ The cluster field from panel~a (inside the box) shown in a larger
context on the POSS~II-R image. 
}
\label{NIRid1}
\end{figure*}

To confirm that WB89-789 is associated with a clustering of stars, in 
Fig.~\ref{onoffcomp} we compare the density of stars in the WB89-789 field 
with that at a nearby off-position (see Sect.~\ref{obsred}). Stars were
counted in concentric rings of equal area ($\pi \times 15 ({\rm pixels})^2$).
For the WB89-789 field the rings were centered on the nebulous region visible 
in Fig.~\ref{NIRid1}, and the counting was performed in the boxed-in area of
the K-frame shown in Fig.~\ref{koverview}a. 
There is a clear excess of stars compared to the
background out to about 22\arcsec\ from the center of the nebulosity; this 
corresponds to a radius of 1.3~pc, in which the average stellar surface 
density is 12~pc$^{-2}$ (compared to $\leq$3~pc$^{-2}$ for the OFF-field).
There is a particularly strong concentration of stars in the inner 15 pixels 
($\sim$ 7\pas5); in this region, with $r \sim 0.4$~pc, the stellar surface 
density is 33~pc$^{-2}$. 

\begin{figure*}[tp]
\resizebox{15cm}{!}{\rotatebox{270}{\includegraphics{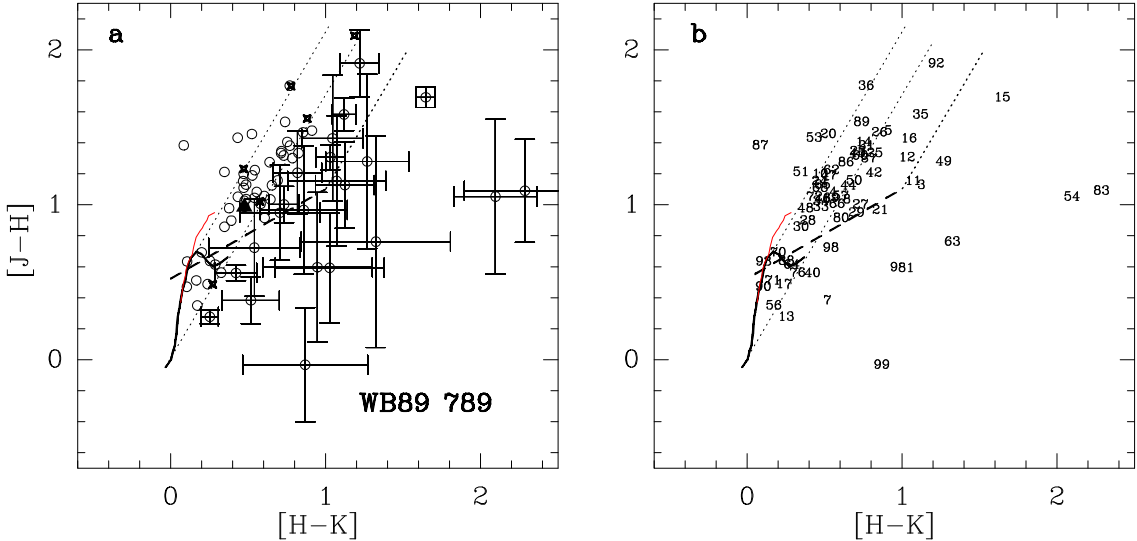}}}
\caption[]{{\bf a.}\ NIR color-color diagram for the 68 stars in 
Fig.~\ref{NIRid1}a
that have reliable detections in all three (J, H, and K) bands. The thick and 
thin drawn curves are the unreddened main sequence and giant branch, 
respectively, from Bessell \& Brett (\cite{bessell}); the 
dashed line indicates the loci of the classical T~Tauri stars (Meyer et
al.~\cite{meyer}); the dotted lines indicate the direction of normal
interstellar reddening (Rieke \& Lebofsky~\cite{rieke}). Crosses on these 
lines mark increments of 5 magnitudes 
of visual extinction from the points where they intersect the main sequence
curve. Objects outside and to the right of the reddening band have intrinsic
NIR excess; to avoid confusion, error bars are shown only for objects that are
well outside the normal reddening band. The filled triangle identifies 
star~33 (see text and Fig.~\ref{NIRid1}). {\bf b.}\ Like a, but with 
the stars identified by their ID-number.}
\label{nirexcess}
\end{figure*}

The central 0\pam9 $\times$ 1\pam1 
part of the region shown in 
Fig.~\ref{koverview}b is where all 5 individual ON-source images overlap; a
(K-frame) contour plot of this region is shown in Fig~\ref{NIRid1}. 
Sixty-eight stars were detected in all three (J, H, K) bands. The photometric 
data are collected in Table~\ref{photomdata}. 
The (J--H), (H--K)-diagram
of these stars is shown in Fig.~\ref{nirexcess}. The dotted lines define the
reddening band for normal stellar photospheres 
(Rieke \& Lebofsky~\cite{rieke}). 
Objects outside and to the 
right of this band have intrinsic NIR excess, 
and for a number of them their location in this diagram can be explained by 
the presence of circumstellar disks, with central holes of various dimensions
(see Lada \& Adams~\cite{ladadams}); these are likely pre-ms stars. 

There are 4 stars outside the normal reddening region on the left 
(nrs.~87, 53, 51, and 20).
All of these except nr.~87 are consistent with a location inside the normal 
region, considering error bars. On the other hand, stars with effective 
temperatures between 2500~K and 4000~K are predicted to have an excess of 
emission in the H-band, which is due to the formation of H$_2$ and absorption
by H$^-$ (Gingerich \& Kumar~\cite{gingerich}); this will shift stars like 
nr.~87 to the left of the normal reddening region in Fig.~\ref{nirexcess} 
(see also Chini et al.~\cite{chini}). 
There are 22 stars outside the normal region on
the right. Eight (nrs.~7, 9, 13, 40, 63, 81, 98, 99) have anomalous colors, 
in that they 
are also below the unreddened TTau locus; these are indicated with circles
in Fig.~\ref{KoverlayC18O}. Three of these (nrs.~13, 40, 98) could be
brought into normal region considering their error bars. The other 5 have
'anomalous' colors, which could be due to unresolved binaries 
(Lada et al.~\cite{lada00}). That leaves at least 14 (22-8) stars with 'true' 
NIR 
excess (i.e., 21\% of the total of 68); these are indicated with a star in
Fig.~\ref{KoverlayC18O}. Eleven are in the Class~II-source region
(between the right-hand normal reddening line and the right-most TTau reddening
line). Their SEDs are dominated by disk emission and their location can
be explained by variation in size of central hole and inclination (see
Lada \& Adams~\cite{ladadams}). Three (nrs.~15, 54, 83) are in the region of 
Class~I sources. These
are the most embedded and youngest objects. Their SED is dominated by emission
from an envelope of gas and dust. 

\subsubsection{Extinction}

\begin{figure*}[tp]
\resizebox{15cm}{!}{\rotatebox{270}
{\includegraphics{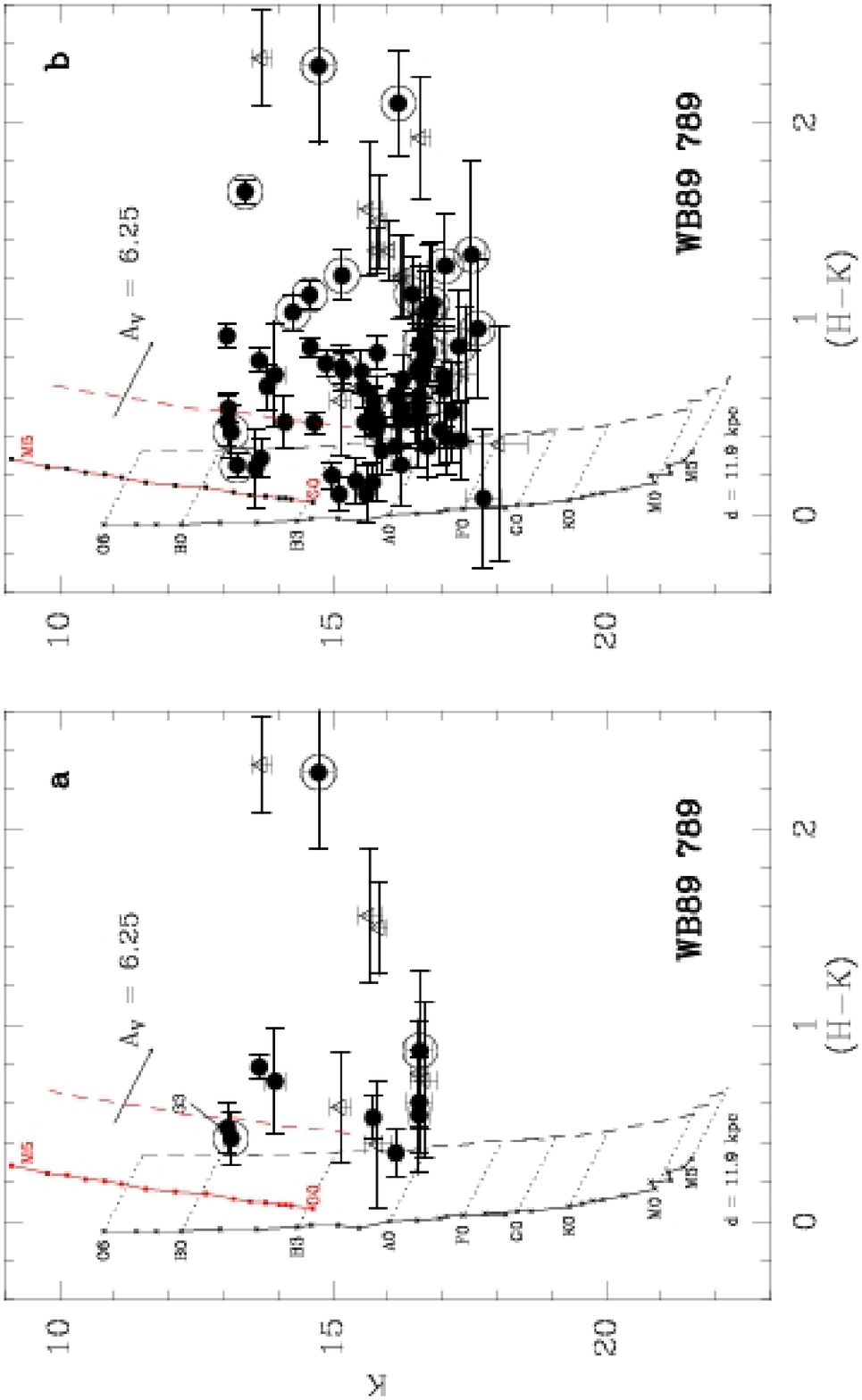}}}
\caption[]{K versus (H-K) color-magnitude diagram for WB89-789.
{\bf a.}\ Only stars within 15 pixels ($\sim$7\pas5) from the center of
the nebula (Fig.~\ref{koverview}b). Filled circles are stars detected in all
three bands (J, H, and K); those with a NIR-excess have been additionally 
marked with an
open circle. Open triangles: stars detected in H and K only. The drawn line
marked O6-M5 is the unreddened main sequence for a distance of 11.9~kpc; data 
from Koornneef (\cite{koornneef}; O6 -- B7) and Bessell \& Brett 
(\cite{bessell}; B8 -- M5). K-magnitudes were derived from M$_V$ 
(Schmidt-Kaler~\cite{schmidt}) and (V-K)$_0$. 
The dashed line is the main sequence for a distance of 11.9~kpc and
a visual extinction A$_{\rm V}$ = 6.25~mags (Rieke \& Lebofsky~\cite{rieke}).
The drawn line marked G0-M5 is the unreddened giant branch (data from Bessell 
\& Brett \cite{bessell}) for a distance of 11.9~kpc. Its location for 
A$_{\rm V}$ = 6.25~mags is shown as a dashed line parallel to the unreddened 
sequence. Star~33 
(see text and Fig.~\ref{NIRid1}) is indicated.
{\bf b.}\ Like a, but for all stars.}
\label{colmag}
\end{figure*}

The amount of visual extinction towards WB89-789 is derived by means of a
K versus (H-K) color-magnitude diagram (Fig.~\ref{colmag}a), plotting only
those stars that are within 7\pas5 from the center of the nebulosity and thus
presumably part of the cluster. Here we can also use those (7) stars
that were detected only in H and K, but not in J. In addition to the stars,
we have drawn the location of the unreddened main sequence for a distance of
11.9~kpc. Ignoring the stars that have been identified as having a NIR-excess
in Fig.~\ref{nirexcess}, because they do not follow the standard interstellar
reddening law, we shift the main sequence along the direction of the
interstellar reddening (indicated by the arrow in Fig.~\ref{colmag}: 
A$_{\rm K} = 0.112 \times {\rm A_V}$ and E(H--K) = 0.061$\times {\rm A_V}$; 
Rieke \& Lebofsky~\cite{rieke}) until it reaches the locations
of the cluster stars. Thus 
we derive a maximum amount of foreground visual extinction of 
A$_{\rm V} \approx 6.25$~magnitudes (maximum, because even the least-reddened
stars can have some amount of internal extinction, i.e., due to the dust in 
the molecular cloud in which they are embedded). This may not seem 
much for such a supposedly distant object, but it is consistent with the
findings of, e.g., 
Fich (\cite{fich87}) and Am\^ores \& L\'epine (\cite{amores}), who concluded 
that the
extinction is fairly low ($\sim 3$~mags.) in much of the outer Galaxy, and
that there does not appear to be a progressively increasing extinction beyond a
few kpc from the Sun. \hfill\break\noindent
In Fig.~\ref{colmag}b we show all 68 stars for which
we have J, H, and K, as well as the 15 stars for which only H and K are
available. Again ignoring stars with NIR-excess (identified with a large open
circle), as well as the stars for which we only have H and K (for which we
cannot say whether they have NIR-excess) we see that there are stars with an
additional extinction of up to $\sim 30$~magnitudes, although the bulk has 
A$_{\rm V} \lsim 7.5$. This is about the same
for stars inside and outside the above-defined cluster radius. 
As we mentioned in the Introduction, the edge of the molecular disk of the 
Galaxy lies at about $R\approx 20$~kpc, thus it is 
unlikely that in our images of WB89-789 there are any stars more distant than 
the cluster, and we conclude that most stars shown are part of the cluster.
Potentially the earliest-type star in the cluster that does not have NIR 
excess is star~33 (see Fig~\ref{NIRid1}): 
If one were to simply shift it back to the the main sequence, it would be 
of spectral type B0, consistent with the spectral type derived from the \Lfir\ 
(and assuming it is an ms-star). 

\begin{figure}
\resizebox{8cm}{!}{\rotatebox{270}
{\includegraphics{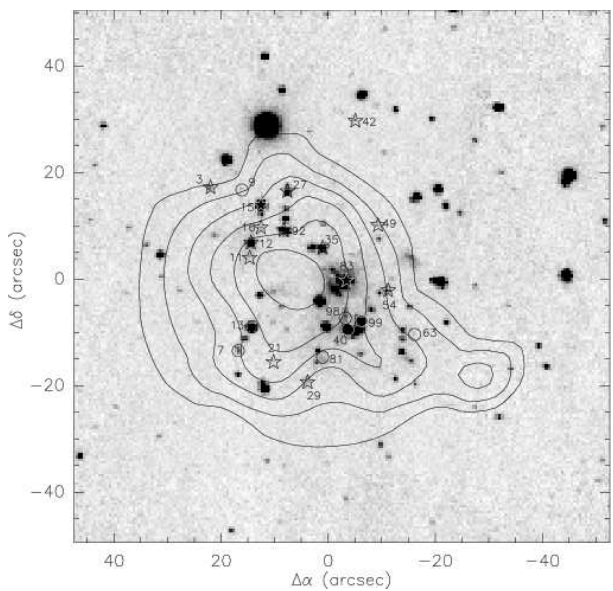}}}
\caption[]{K-band image of the WB89-789 region, with contours of the integrated
C$^{18}$O(2--1) emission superimposed (see Fig.~\ref{comap}c for details).
The IRAS position is at (0,0). 
The stars with NIR-excess are labeled, and identified with asterisks; those 
with anomalous colors (see text) are marked with circles. 
}
\label{KoverlayC18O}
\end{figure}

\begin{figure*}[tp]
\resizebox{15cm}{!}{\rotatebox{270}{
\includegraphics{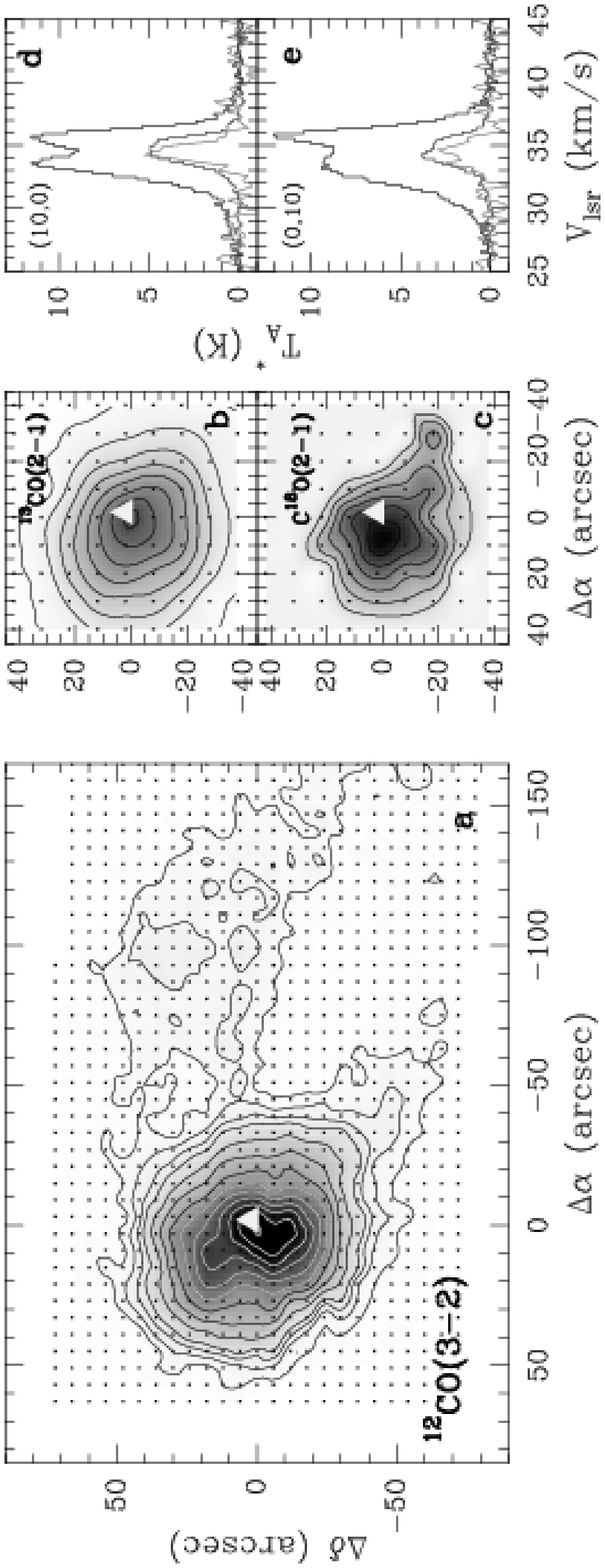}}}
\caption[]{JCMT-data. {\bf a.}\ Map of the integrated intensity of CO(3--2) 
associated with WB89-789 ($\int T^*_{\rm A}{\rm d}v$; the integral is over the 
bulk of the emission, $30< V_{\rm lsr} <38$~\kms, and excludes the line 
wings). 
Contour levels 2.5(2.5)10(5)55~K\kms\ (lowest(step)highest).
The filled triangle marks the peak of the SCUBA 
850~$\mu$m emission (at $-$1\pas9, 2\pas1), and the observed positions are 
indicated with the tiny crosses. The IRAS point source position is at offset 
(0,0).
{\bf b.}\ Map of the $^{13}$CO(2--1) emission (area under the fitted 
Gaussians). 
Contour levels are 1(2)15~K\kms. Triangle and crosses as in a. {\bf c.}\ As b, 
but for the C$^{18}$O(2--1) emission. Contour levels are 0.3(0.3)2.1~K\kms.
{\bf d.}\ Line profiles at offset position (10\arcsec,0\arcsec). The strongest
emission is the $^{12}$CO(2--1)-line; the profile clearly shows 
self-absorption. The less intense emission, plotted as a histogram, is the 
$^{13}$CO(2--1) line, and the drawn line is that of C$^{18}$O(2--1), multiplied
by 4. {\bf e.}\ Like d, but for offset position (0\arcsec, 10\arcsec).}
\label{comap}
\end{figure*}

Figure~\ref{KoverlayC18O} shows a section of the K-band image of the 
WB89-789 area, with the contours of integrated C$^{18}$O(2--1) emission (see
Fig.~\ref{comap}c) superimposed.  
The 14 stars with NIR-excess are shown in Fig.~\ref{nirexcess}b, and 
the 8 stars with anomalous colors are marked in 
Fig.~\ref{KoverlayC18O}. Rather than being concentrated
in the center of the cluster (which we take to be the nebulous region seen
in Fig.~\ref{NIRid1}), the NIR-excess stars are distributed in a ring around
the peak of the C$^{18}$O(2--1) emission. At the location of the molecular
peak, no stars are seen in this K-image, which may be because the extinction
there is too high. The average column density of H$_2$, N(H$_2$), derived from 
the C$^{18}$O(2--1) data (see Sect.~3.3.1) is 7.8$\times 10^{21}$~cm$^{-2}$. 
Because N(H$_2$)=$10^{21} \times$A$_{\rm V}$ (Bohlin et al.~\cite{bohlin}, and
assuming the gas-to-dust ratio in the FOG is the same as locally) 
this implies a visual extinction A$_{\rm V}$ of about 8~magnitudes due to the
whole of the molecular clump (front to back). At the
peak of the C$^{18}$O(2--1) emission 
N(H$_2$)$\approx 1.0 \times 10^{22}$~cm$^{-2}$, and thus A$_{\rm V} \approx 
10$~magnitudes. If the cluster stars are embedded in the molecular gas, as 
they are likely to be, then the general (i.e., not counting that which is due
to the circumstellar material) internal extinction would therefore be on 
average $\sim$5~magnitudes (and in any case $\leq$10~magnitudes), consistent 
with the number derived above from the color-magnitude diagram.

\subsection{Molecular cloud}

\subsubsection{Cloud parameters}

\smallskip\noindent
{\it Whole cloud}

BW94 mapped the molecular cloud associated with WB89-789 in CO(1--0) with
the SEST (beam size 45\arcsec), on a 40\arcsec\ raster. We have obtained 
new, better sampled CO(2--1) and (3--2) maps with the JCMT; the latter is 
shown in Fig.~\ref{comap}a. 
From the CO(2--1) data, following the procedures outlined in BW94,
we derive a mass $M_{\rm CO} \approx 4.5 \times 10^3$~\Msol\ 
(from N(H$_2$)=$X$ $\int T^*_{\rm R}(CO)dv$, with $X=1.9 \times
10^{20}$~cm$^{-2}$(K\kms)$^{-1}$ and using $T_{\rm mb}$=
$\int T^*_{\rm A}/\eta_{\rm fss}{\rm d}v$; $\eta_{\rm fss} = 0.80$ is the 
forward efficiency). The virial mass 
$M_{\rm vir} \approx 5.1 \times 10^3$~\Msol\ 
($M_{\rm vir}=126 r_{\rm e} (\Delta V)^2$ for an $r^{-2}$ density
distribution; MacLaren et al.~\cite{maclaren}); $\Delta V$ is the FWHM line 
width, and the equivalent, 
beam-corrected radius $r_{\rm e} \approx 5.3$~pc. These numbers 
agree very well with those derived by BW94 from the CO(1--0) map. We also
note that the virial mass and the mass derived from the empirical method,
using $X$, are in excellent agreement. The value of $X$ quoted above is the 
inner Galaxy value; Brand \& Wouterloot (\cite{bw95}) have shown that in the 
outer Galaxy $X$ may be somewhat larger, but lies within $30-45$\% of the 
inner Galaxy value. Hence at the edge of the Galaxy, $X$ may be 
$2.8 \times 10^{20}$~cm$^{-2}$(K\kms)$^{-1}$, which results in 
$M_{\rm CO} \approx 4.5-6.6 \times 10^3$~\Msol. These results are collected in 
Table~\ref{cloudlineparams}.

\smallskip\noindent
{\it Cloud core: {\bf CO}}

\noindent
In Fig.~\ref{comap}b and c we show a plot of the integrated emission of 
$^{13}$CO(2--1) and C$^{18}$O(2--1), respectively. The emission in these two
isotopomers was mapped only in the core of the cloud, but as can be seen
from the maps, at least for C$^{18}$O(2--1), all the emission is 
contained within 
the mapped region. While the $^{13}$CO and $^{12}$CO(2--1) emission peak at 
the IRAS position (offset 0\arcsec,0\arcsec), the C$^{18}$O(2--1)-peak is  
offset by about 10\arcsec\ to the East; this is not due to pointing 
uncertainties because $^{13}$CO and C$^{18}$O were observed simultaneously.
The dust continuum emission (see Fig.~\ref{cont}) also peaks very close to the 
IRAS position; hence the offset detected in the C$^{18}$O emission may be 
due to that molecule's depletion near the core in which the IRAS source is
embedded (cf. Tafalla et al.~\cite{tafalla}). The $^{12}$CO emission, being 
more optically thick, only traces the outer layers 
of the core; also, $^{13}$CO is more optically thick 
than C$^{18}$O (see later) and traces the more outlying layers. 
The C$^{18}$O(2--1) core is elongated in a
SW-direction, similar to the $^{12}$CO core in panel a. 
The core-radius derived from the $^{13}$CO observations is identical to that,
derived from $^{12}$CO, while the C$^{18}$O core is $\sim$40\% smaller.
The masses $M_{\rm CO}$ and $M_{\rm vir}$ determined from $^{12}$CO in the 
same region as mapped in $^{13}$CO and C$^{18}$O are reported in Cols.~4 and 6,
respectively, of Table~\ref{cloudlineparams}, under ``$^{12}$CO, core''.

\begin{table}
\caption{Cloud parameters derived from the molecular line measurements.}
\label{cloudlineparams}
\begin{flushleft}
\begin{tabular}{l|cc|l|ccl}
\hline\noalign{\smallskip}
\multicolumn{1}{c|}{} & \multicolumn{1}{c}{$r_{\rm e}^1$} 
 & \multicolumn{1}{c|}{$r_{\rm hp}^1$}
 & \multicolumn{1}{c|}{$\Delta V$}
 & \multicolumn{1}{c}{$M_{\rm CO}^2$} & \multicolumn{1}{c}{$M_{\rm lte}^3$} 
 & \multicolumn{1}{c}{$M_{\rm vir}^4$} \\ 
\multicolumn{1}{c|}{} & \multicolumn{2}{c|}{pc} 
 & \multicolumn{1}{c|}{km\,s$^{-1}$}
 & \multicolumn{3}{c}{$10^3$~\Msol} \\
\hline\noalign{\smallskip}
 $^{12}$CO, cloud & 5.3 & 1.6 & 2.77$^5$     & 4.5--6.6  &     & 5.1 \\
                  &     &     & 2.69$^{5,6}$ &      &          & 4.8$^6$  \\
 $^{12}$CO, core  & 2.9 & 1.6 & 3.34$^5$     & 3.2--4.7  &     & 4.1  \\
                  &     &     & 3.22$^{5,6}$ &      &          & 3.8$^6$  \\
 $^{13}$CO, core  & 2.7 & 1.1 & 2.14         &      & 1.0--5.0 & 1.6      \\
 C$^{18}$O, core  & 1.6 & 0.9 & 1.72         &      & 0.8--3.2 & 0.6      \\
 CS, core         & 1.1 & 0.7 & 2.07         &      &          & 0.4      \\
\noalign{\smallskip}
\hline
\end{tabular}

\smallskip\noindent
 \\
1.\ $r_{\rm e}$: Equivalent radius, corrected for beam size, at the 
$2-3\sigma$-level; $r_{\rm hp}$: the radius at half-maximum intensity level, 
corrected for beam size. \\
2.\ Via N(H$_2$)=$X$ $\int T_{\rm mb}(CO)dv$.
The indicated range of masses
derives from using the inner Galaxy value 
$X=1.9 \times 10^{20}$~cm$^{-2}$(K\kms)$^{-1}$ and allowing a possible 45\% 
increase at the edge of the Galaxy (see text).\\
3.\ Assuming LTE-conditions, and $T_{\rm ex}$=20~K. Smaller values are for 
local abundance ratios for [H$_2$]/[$^{13}$CO] and [H$_2$]/[C$^{18}$O], higher
values are for abundances extrapolated to the FOG (see text). \\
4.\ $M_{\rm vir}$=126 $r$ ($\Delta V)^2$, for a density distribution 
$\propto r^{-2}$, and using $r_e$. \\
5.\ Equivalent width, derived from the area under the 
emission line and the peak temperature.\\
6.\ Excluding the wing emission. \\
\end{flushleft}
\end{table}

Column densities $N$ can be derived from $T_{\rm mb}(^{13}$CO) and 
$T_{\rm mb}$(C$^{18}$O), assuming the emission is optically thin 
(see, e.g., Rohlfs \& Wilson~\cite{rohlfs}; Brand \&
Wouterloot~\cite{s151}, Eq.~1). The excitation temperature \Tex\ is 
derived from the optically thick $^{12}$CO transition.
The $^{12}$CO emission line is not just optically thick:
the line profiles within 20--30\arcsec\ from the IRAS position 
show self-absorption. Examples are shown in Fig.~\ref{comap}d, e; we
shall return to this in Sect.~3.2.2.
From the emission at positions away from the region with self-absorption, we
find that \Tex $\sim 20$~K is a representative value for this cloud. 
At each grid position of the cloud-map we derived N($^{13}$CO) and 
N(C$^{18}$O).
Column densities of H$_2$, and LTE-masses $M_{\rm lte}$ were then 
derived through N(H$_2$) $\approx 5\times 10^5$N($^{13}$CO) (Dickman \&
Clemens~\cite{dickman}) and 
N(H$_2$) $\approx 6\times 10^6$N(C$^{18}$O) (Frerking et al.~\cite{frerking}). 
The $M_{\rm lte}$ 
are presented in Col.~5 of Table~\ref{cloudlineparams}.
From the detection equation, and the adopted value of \Tex, we find that on 
average $\tau$(C$^{18}$O)
$\sim 0.05$ (peak value 0.11 at offset 10\arcsec,0\arcsec), and
$\tau$($^{13}$CO) $\sim 0.22$ (peak value 0.64 at offset 
0\arcsec,$-$10\arcsec). 
It is clear that the C$^{18}$O is optically thin,
whereas the $^{13}$CO is less so. 
\hfill\break\noindent
The $M_{\rm lte}$
derived from $^{13}$CO and C$^{18}$O are in good agreement, but they are both 
smaller than the core-masses derived from $^{12}$CO. This is partly due to
the fact that in $^{12}$CO one sees the outer layers of the cloud,
not seen in the other two isotopomers.
In the derivation of the LTE-masses we have used the locally-determined
abundance ratios [H$_2$]/[$^{13}$CO] and [H$_2$]/[C$^{18}$O]. However, 
because of
the galactic abundance gradients these values are certainly not valid at
the edge of the Galaxy. As argued by Brand \& Wouterloot (\cite{bw95}), the
ratio [H$_2$]/[$^{13}$CO] in the far-outer Galaxy might be about 5 times 
higher than locally. A similar extrapolation of the abundance gradients 
(Wilson \& Matteucci~\cite{wilson}; Wilson \& Rood~\cite{rood}) suggests that 
the ratio [H$_2$]/[C$^{18}$O] in the far-outer Galaxy might be about 4 times 
higher than the local value.
Taking this into account, $M_{\rm lte}$ derived from $^{13}$CO and C$^{18}$O 
would become $5 \times 10^3$~\Msol\ and $3.2 \times 10^3$~\Msol, respectively, 
which is closer to 
the range for $M_{\rm CO}$ (see Table~\ref{cloudlineparams}).

\smallskip\noindent
{\it Cloud core: {\bf CS}}

\noindent
All CS lines are detected essentially only at the IRAS point source position.
The lines are Gaussian, showing no signs of outflow or
other asymmetries. From the CS(2--1) map we find a (beam-corrected) radius of
the CS-core of 0.7~pc, smaller even than the C$^{18}$O core. With an average
line width $\Delta V \approx 2.07$~km\,s$^{-1}$, we derive a virial mass of
$M_{\rm vir} \approx 400$~\Msol, similar to what was found from the C$^{18}$O
data.
We have convolved the CS(3--2) and (5--4) spectra at the central position to
the CS(2--1) beam, assuming Gaussian beam profiles. The resulting spectra 
were put on a main-beam brightness temperature scale ($T_{\rm mb}$), and
used in an LVG-model to derive density and kinetic temperature of the CS-gas.
We ran the model with a range of parameters: T=10--100~K in steps of 2~K, 
log(n$_{\rm H_2}$)=4--6 in steps of 0.1--0.2, and abundances 1$\times 10^{-11},
3 \times 10^{-11}, 5 \times 10^{-11}$, and $1\times 10^{-10}$. We used
a velocity gradient $dv/dr \sim 1.5$~km\,s$^{-1}$pc$^{-1}$.
We found the smallest $\chi^2$ ($\sim 0.6$), hence the best reproduction of 
the observed line ratios, for a CS-abundance of 3 $\times 10^{-11}$, 
$T_{\rm kin} \approx 84$~K, and log(n$_{\rm H_2}/{\rm cm}^{-3}) 
\approx 4.9$. The fit is not very sensitive to the temperature, as $\chi^2$ is
within a factor of 2 of the minimum value over a range of about 30~K. 
Nevertheless it is clear that 
the deeper layers of the core from which the CS emission 
originates are considerably hotter than the outer parts traced by 
$^{12}$CO.

\subsubsection{Molecular outflow}

\begin{figure*}[tp]
\resizebox{15cm}{!}{\rotatebox{270}{\includegraphics{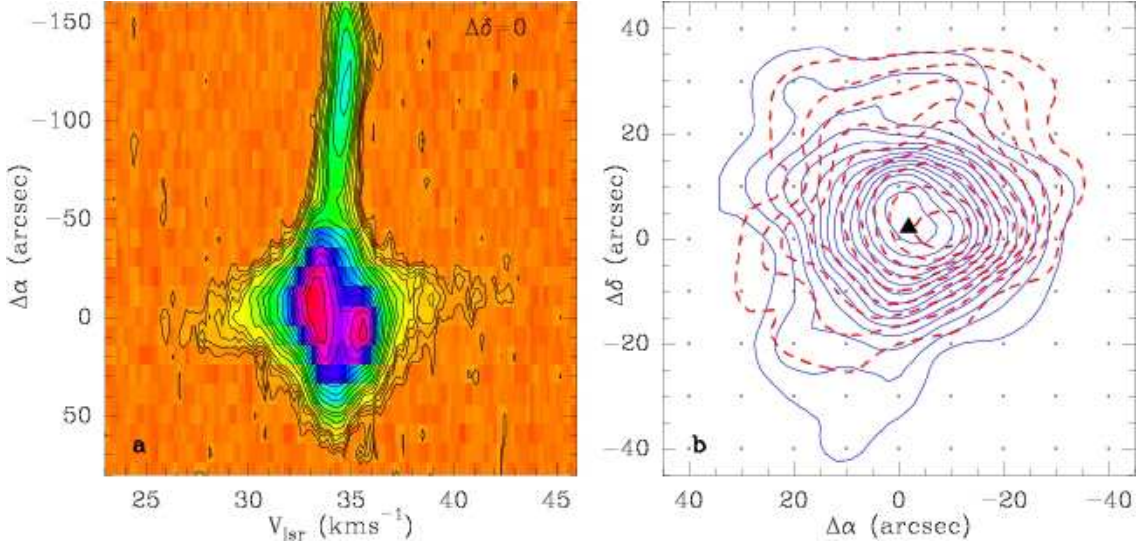}}}
\caption[]{{\bf a.}\ $\Delta \alpha$ -- velocity diagram at
$\Delta \delta$=0\arcsec\ of the $^{12}$CO(2--1) emission. Contour levels are
0.25, 0.4(0.2)1(0.5)3(1)15~K. At the location of the IRAS point source (at 0,0)
emission
in the wings of the line profiles is evident. {\bf b.}\ Map of the 
integrated wing emission. Blue (drawn): $24.7< V_{\rm lsr} <31.7$~\kms; 
red (dashed): $36.7< V_{\rm lsr} <43.7$~\kms. Contour levels and step are
0.5~K\kms.
The filled triangle marks the position of the peak of the SCUBA 850~$\mu$m 
emission ($-$1\pas9, 2\pas1); the observed positions are indicated with the 
crosses.}
\label{wings}
\end{figure*}

The SEST CO(1--0) spectrum at the IRAS position showed a line profile with
non-Gaussian wings (BW94). The new, higher-resolution CO(2--1) data allow us 
to better
outline the outflow (see Fig.~\ref{wings}a), which is completely
contained within one SEST CO(1--0) beam ($\sim 45$\arcsec; Fig.~\ref{wings}b). 
%
%
In Fig.~\ref{wings}a we show a position-velocity diagram at 
$\Delta\delta$=0\arcsec. Enhanced emission in the line wings is clearly 
visible at $\Delta\alpha$=0\arcsec. 
Also clearly visible is a double-peak structure in the 
$^{12}$CO(2--1) emission. At positions within $\sim 20$\arcsec\ from the 
center of
the map the lines show a significant indentation, as illustrated in 
Fig.~\ref{comap}d, e. By comparing the position-velocity map in 
Fig.~\ref{wings}a with the theoretical contour maps shown in Figs.~3 and 4 
in Phillips et al. (\cite{phillips}), we conclude that the pattern seen in 
Fig.~\ref{wings}a is due to self-absorption (i.e., by a temperature gradient in
the cloud core), rather than due to absorption by a colder (unrelated) 
foreground cloud.
This is consistent with the fact that the CS-data reveal a higher temperature
than the $^{12}$CO (Sect.~3.2.1).

Again assuming LTE-conditions, we calculated the mass of the gas in the line
wings from the $^{12}$CO(2--1) data; the wings are not visible 
in the other two isotopomers. We 
adopted \Tex = 20~K, used $T_{\rm mb}$
(see Sect.~3.2.1), and considered 
both the optically thick and -thin cases. 
The total velocity intervals, $24.7-31.7$~\kms\ (blue wing) and 
$36.7-43.7$~\kms\ (red wing), over which was integrated (in 1~\kms-wide bins), 
were chosen such that the emission of the quiescent gas was excluded. 

In addition to the mass, we also calculated the flow's energy, momentum, size, 
and velocity (as outlined, e.g., by Lada~\cite{lada}; Wouterloot \& 
Brand~\cite{vela}); the dynamical 
age was then derived from the ratio of the flow's size ($\sim 3.5$~pc) and its 
velocity. The outflow parameters are collected in Table~\ref{flowparams}.
Here we also give the flow's mechanical luminosity ($L_{\rm m}$=
Energy/age) and the force needed to drive the flow ($F_{\rm m}$= Momentum/age).

\begin{table*}
\caption{Parameters of the molecular outflow associated with
WB89-789$^{\dagger}$}
\label{flowparams}
\begin{flushleft}
\begin{tabular}{l|ccc|ccc|ccc|c|c|c|ccc|ccc}
\multicolumn{19}{l}{{\bf CO(2--1)}} \\
\hline\noalign{\smallskip}
\multicolumn{1}{l|}{Wing} 
 & \multicolumn{3}{c|}{Mass}
 & \multicolumn{3}{c|}{Mom.}
 & \multicolumn{3}{c|}{Energy} 
 & \multicolumn{1}{c|}{Size}
 & \multicolumn{1}{c|}{$V_{\rm out}$}
 & \multicolumn{1}{c|}{$t_{\rm dyn}$} 
 & \multicolumn{3}{c|}{$F_{\rm m}$} & \multicolumn{3}{c}{$L_{\rm m}$} \\
\multicolumn{1}{l|}{} 
 & \multicolumn{3}{c|}{\Msol}
 & \multicolumn{3}{c|}{\Msol\ km\,s$^{-1}$} 
 & \multicolumn{3}{c|}{$10^{44}$~erg}
 & \multicolumn{1}{c|}{pc} 
 & \multicolumn{1}{c|}{km\,s$^{-1}$}
 & \multicolumn{1}{c|}{$10^5$~yr}
 & \multicolumn{3}{c|}{10$^{-4}$~\Msol\ km\,s$^{-1}$\,yr$^{-1}$} 
 & \multicolumn{3}{c}{10$^{-2}$~\Lsol} \\
\hline\noalign{\smallskip}
blue & 5.1 & 3.2 & 4.1 & 18 & 11 & 15 & 6.7 & 4.2 & 5.4 & 1.4 & 3.5 & 4.0 & 0.5 & 0.3 & 0.4 & 1.4 & 0.9 & 1.1 \\
red  & 4.0 & 2.6 & 3.3 & 14 &  9 & 11 & 4.7 & 3.0 & 3.8 & 1.6 & 3.3 & 4.8 & 0.3 & 0.2 & 0.2 & 0.8 & 0.5 & 0.7 \\
\noalign{\smallskip}
\hline
\noalign{\smallskip}
\multicolumn{19}{l}{{\bf CO(3--2)}} \\
\hline\noalign{\smallskip}
blue & 6.4 & 4.0 & 3.4 & 28 & 18 & 15 & 15 & 10 &  8 & 1.1 & 4.4 & 2.4 & 1.2 & 0.7 & 0.6 & 5.2 & 3.3 & 2.8 \\
red  & 5.2 & 3.3 & 2.8 & 27 & 17 & 14 & 20 & 13 & 11 & 1.2 & 5.1 & 2.3 & 1.2 & 0.7 & 0.6 & 7.2 & 4.6 & 3.9 \\
\noalign{\smallskip}
\hline
\noalign{\smallskip}
\multicolumn{19}{l}{${\dagger}$\ In case of three numbers per column, 
left: $T_{\rm ex}=20$, $\tau = 1$, center: $T_{\rm ex}=20$, $\tau = 0$, 
right: $T_{\rm ex}=40$, $\tau = 0$.} \\ 
\multicolumn{19}{l}{We used the local abundance ratio [CO]/[H$_2$] = 
1.0 $\times 10^{-4}$.} \\
\end{tabular}
\end{flushleft}
\end{table*}

\begin{figure*}[tp]
\resizebox{15cm}{!}
{\includegraphics{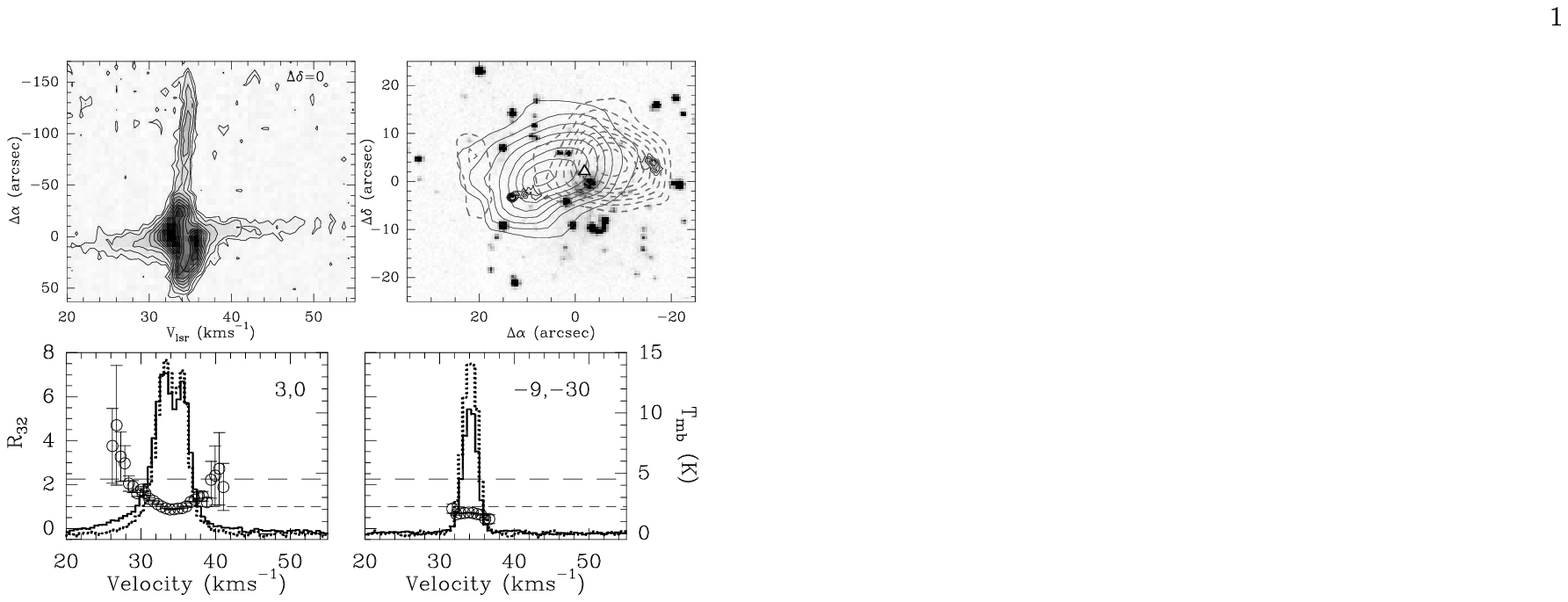}}
\caption[]{{\bf Top.}\ (Left) $\Delta \alpha$ -- velocity diagram at
$\Delta \delta$=0\arcsec\ of the $^{12}$CO(3--2) emission. Contour levels are
0.5, 1(1)15~K.
(Right) Map of the integrated $^{12}$CO(3--2) wing emission. 
Blue (drawn): $20< V_{\rm lsr} <31.2$~\kms; 
red (dashed): $37.2< V_{\rm lsr} <55$~\kms. Contour levels and step are
2.0~K\kms.
The IRAS position is at (0,0).
The white triangle with black outline marks the position of the peak of the 
SCUBA 850~$\mu$m emission.
The underlying image is the K-frame of the region. Contours are drawn to 
highlight two extended features near offsets ($+$10\arcsec,$-$3\arcsec) and 
($-$16\arcsec,4\arcsec) that might mark spots where the outflow impacts with 
the ambient medium.
{\bf Bottom.}\ Line profiles and ratios at two positions: near the 
center of the flow and the peak of the blue lobe (left) and in a quiescent 
part of the cloud (right). The offsets are shown inside the panels. The drawn
line profile is that of CO(3--2), the dotted one if for CO(2--1); the 
temperature-scale is shown on the right-hand axis. The open circles with error
bars indicate the line ratio $R32$ for those velocity channels where 
$T_{\rm mb} > 1.5\sigma$ in both lines; values are shown on the 
left-hand
axis. Short-dashed and long-dashed lines mark the upper limit of the line 
ratio if both lines are optically thick or thin, respectively (see text).}
\label{R32flow}
\end{figure*}

\smallskip
Our $^{12}$CO(3--2) data, taken with a smaller beam size and smaller 
grid spacing, show self-absorbed profiles in the cloud core. Line wings 
are more extended than for CO(2--1), and reach 20~\kms\ (blue) and 55~\kms\ 
(red), respectively. The top panels of Fig.~\ref{R32flow} show a 
position-velocity diagram at $\Delta\delta$=0\arcsec\ (left), and a contour 
plot of the outflow lobes in the CO(3--2)-emission (right).

We have convolved the CO(3--2) observations with a Gaussian beam to the 
resolution of the CO(2--1) data, and resampled all (2--1) and (3--2) spectra
to a velocity resolution of 0.55~\kms. Using $\eta_{\rm fss} = 0.71, 0.80$ for
CO(3--2, 2--1), we have determined the line ratio 
$R32 = T_{\rm mb}[{\rm CO}(3-2)]/T_{\rm mb}[{\rm CO}(2-1)]$
in each velocity channel where $T_{\rm mb} > 1.5\sigma$ for both lines.
For optically thick lines the ratio $R32$ tends to 1 
for large values of $T_{\rm kin}$, while the limit for optically thin lines 
is 2.25 (Bachiller \& Tafalla~\cite{bachiller}; Codella et al.~\cite{codella}).
In the lower panels of Fig.~\ref{R32flow} we show the
results at two representative positions: near the center of the flow and the 
peak of the blue lobe (lower left) and at a more quiescent location in the 
cloud (lower right). Typical values for $R32$ in the line center are 
0.80--0.90, corresponding to $T_{\rm kin} \approx 15-30$~K, assuming both 
lines are optically thick; this is consistent with what we found from the 
CO(2--1) lines, and from a fit to the spectral energy distribution (see 
Sect.~3.2.3). In the line wings, typically $R32 < 1.5$, implying 
$T_{\rm kin} \lsim 40$~K if both lines are optically thin away from the line 
center. But $R32$ can reach values of 4 to 6 at some positions, as can be seen 
in the panel on the lower left in Fig.~\ref{R32flow}. This cannot be explained
if both lines are optically thin or thick at the same time.
On the other hand, the 
uncertainties in the high $R32$-values are large, and assuming a more stringent
criterion in their calculation, e.g., by requiring the value of 
$T_{\rm mb}$
to be at least 3$\sigma$ in both transitions, would obviously remove most
(though not all) points with anomalous values of $R32$. Ratios $R32$ of around
2 are found for optically thin lines and $T_{\rm kin} \approx 120$~K, which
is not unexpected in the shocked gas in molecular outflows (see, e.g., 
Hatchell et al.~\cite{hatchell}; Hirano \& Taniguchi~\cite{hirano}). 
A more accurate value of the temperature of
the outflowing gas, and the possible existence of a gradient along the lobes,
can be obtained only by observing higher transitions and at higher resolution
(the JCMT beam sizes at the frequencies of the CO(3--2) and (2--1) transitions
correspond to 0.8 and 1.2~pc, respectively)

\smallskip
Emission lines in the ``tail'' of the cloud, i.e., at offsets $\Delta\alpha 
\lsim -60$\arcsec\ (see Fig.~\ref{comap}) are less intense than in the core,
with $T_{\rm mb} \le 5$~K and $\le 7$~K for CO(3--2) and (2--1), respectively.
The lines are narrow ($\Delta V_{\rm fwhm} \sim 1.1 - 1.5$~\kms), becoming 
progressively narrower towards the W-edge of the cloud. There is a slight 
velocity gradient in this part of the cloud, as can be seen in 
Figs.~\ref{wings}a and \ref{R32flow} (top, left). Line ratios $R32$ indicate
typical values of $T_{\rm kin} \approx 8 \pm 2$~K in this region.

\smallskip
Following the same procedure as outlined above, we have calculated various 
outflow parameters from the CO(3--2) data; these too are collected in 
Table~\ref{flowparams}. 
The outflow in the CO(3--2) line is stronger and more extended in 
velocity, and more compact than that in the CO(2--1) line. The smaller size and
higher average mass-weighted outflow velocity for the 3--2 flow result in a 
lower dynamical age. Other flow parameters are likewise influenced by this, 
but as can be seen from Table~\ref{flowparams}, in general the results obtained
from the two transitions are consistent within factors of a few. 
The lobes of this outflow have considerable overlap, indicating 
that the outflow axis is pointing close to the line-of-sight. As a 
consequence the size derived here is likely to be a lower limit, and age 
an upper limit. Values of $F_{\rm m}$ and $L_{\rm m}$ are thus lower limits 
too.
\hfill\break\noindent
In the calculation of the flow parameters in Table~\ref{flowparams} we 
have used the local abundance of [CO]/[H$_2$]. It is possible that the 
abundance ratio is lower at larger galactocentric distance. Gradients in 
C/H and O/H are similar (Wilson \& Matteucci~\cite{wilson}; Rolleston et 
al.~\cite{rolleston}), and assuming that the CO-abundance is determined
by that in C, one expects [CO]/[H$_2$] to be a factor of about 4 lower
in the far-outer Galaxy. This would increase all the parameters listed in 
Table~\ref{flowparams} by a factor of 4, except for the size, flow velocity, 
and age. Keeping this in mind, 
a comparison with outflows associated with IR sources of similar luminosity 
(Beuther et al.~\cite{beuther}; Zhang et al.~\cite{zhang}) shows that this is 
not inconsistent with a typical outflow in an intermediate-mass star-forming 
region.

\smallskip
In Fig.~\ref{R32flow}, top right panel, 
we show the CO(3--2) outflow superimposed on the K-image.
Near offsets ($+$10\arcsec,$-$3\arcsec) and ($-$16\arcsec,4\arcsec), bright 
extended spots are seen (cf. also Fig.~\ref{NIRid1}a). 
The spots are brightest in K, and virtually absent in J 
(cf. Fig.~\ref{koverview}b). 
This emission might be due to the 2.1~$\mu$m line of H$_2$, a 
manifestation of the interaction between outflow and ambient medium. 
This can be established only by taking a spectrum.

\begin{table}
\caption{Cloud parameters derived from the continuum measurements.}
\label{cloudcontparams}
\begin{flushleft}
\begin{tabular}{lccrc}
\hline\noalign{\smallskip}
\multicolumn{1}{c}{} & \multicolumn{1}{c}{$r_{\rm e}^1$}
 & \multicolumn{1}{c}{$r_{\rm hp}^1$} 
 & \multicolumn{1}{c}{S$^2$}
 & \multicolumn{1}{c}{$M_{\rm dust}^3$} \\
\multicolumn{1}{c}{} & \multicolumn{2}{c}{pc} 
 & \multicolumn{1}{c}{Jy}
 & \multicolumn{1}{c}{\Msol} \\
\hline\noalign{\smallskip}
 450~$\mu$m       & 0.85 & 0.25 & 12.9 &    \\
 850~$\mu$m       & 1.48 & 0.39 &  1.8 &    \\
 1.2~mm           & 1.50 & 0.76 &  0.5 &    \\
 SED fit          &      &      &      & 11 \\
\noalign{\smallskip}
\hline
\end{tabular}

\smallskip\noindent
 \\
1.\ $r_{\rm e}$, $r_{\rm hp}$: radius at 3$\sigma$-level and half-peak level,
respectively; both corrected for beam size. \\
2.\ Integrated flux density above the 3$\sigma$-level \\
3.\ Assuming 
optically thin dust emission, 
a dust absorption coeff. 1~cm$^2$g$^{-1}$ at 250~GHz; the exponent of the 
freq. dependence of the dust optical depth $\beta=1.8$, and $T_{\rm dust}$=23~K
(see text for details). \\
\end{flushleft}
\end{table}

\subsubsection{Dust continuum}
\begin{figure*}[tp]
\resizebox{15cm}{!}{\rotatebox{270}{
\includegraphics{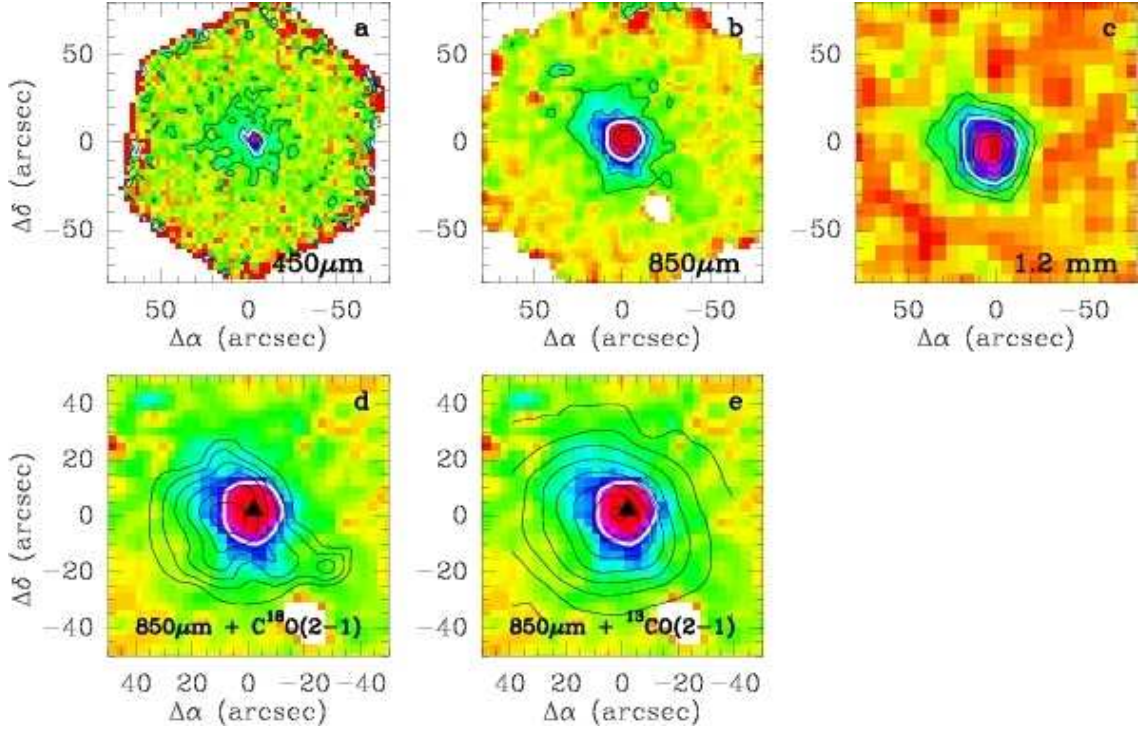}}}
\caption[]{
Images of continuum emission at 450, 850, and 1200~$\mu$m, and overlays of
the 850~$\mu$m emission with the $^{13}$CO and C$^{18}$O emission. The
IRAS position is at (0,0).
{\bf a.}\ SCUBA 450~$\mu$m map. Contour values 1(1)5~Jy/beam. The white 
contour indicates the half-peak level. {\bf b.}\ SCUBA 850~$\mu$m map. 
Contours 0.1(0.1)0.7~Jy/beam. White contour as in a. {\bf c.}\ SIMBA 1.2~mm
map. Contours 45(30)230~mJy/beam. White contour as in a. 
{\bf d.}\ SCUBA 850~$\mu$m map with C$^{18}$O(2--1) contours superimposed. 
Contour values are 0.3(0.3)2.1~K\kms. The white contour indicates the 
850~$\mu$m emission half-peak level, while the triangle marks its peak. 
{\bf e.}\ As d, but with $^{13}$CO(2--1) contours superimposed. Contour values 
1(2)15~K\kms. White contour and triangle as in d.
}
\label{cont}
\end{figure*}

In Fig.~\ref{cont} we have collected the maps of the dust continuum, made at
3 different wavelengths, and shown their location with respect to the core
found in the molecular lines. The $^{13}$CO emission peaks very close to 
the maximum in the 850~$\mu$m map, while the C$^{18}$O peak is offset by about
10\arcsec\ (1 grid point $\sim$ half a beam width) to the East.
Flux densities at the three wavelengths were calculated by integrating all
emission above the 3$\sigma$-contour, and sizes of the dust-emission cores
were determined from the half-peak value contour and were corrected for beam 
size. 
The results are collected in Table~\ref{cloudcontparams}.

\begin{figure}
\resizebox{8cm}{!}{
\includegraphics{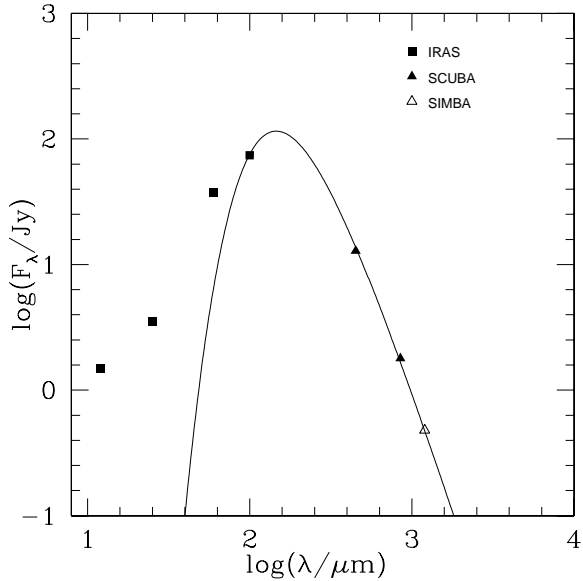}}
\caption[]{
Spectral energy distribution (SED) of WB89-789. Along the axes are log$\lambda$
and log$F$. The data shown are our own 
measurements of the continuum emission at 450, 850, and 1200~$\mu$m, together
with the flux densities in the 4 IRAS bands (IPSC2). The curve shows
the best fit to the data points at wavelengths $\geq 100$~$\mu$m, which 
has $T_{\rm dust}=23$~K, and a dust absorption coefficient of 1.8,  
and which was obtained by assuming 
a dust opacity at 250~GHz of 1~cm$^2$g$^{-1}$.
}
\label{sed}
\end{figure}

Figure~\ref{sed} shows the spectral energy distribution (SED) of WB89-789. We 
have plotted our three FIR and mm-data points, in addition to the 4 IRAS flux
densities, taken from the IPSC2. We have made a grey-body fit to the data 
points for 100~$\mu$m onwards; the fit is shown as the drawn curve. The 12, 
25, and 60~$\mu$m points cannot be fit by the same curve. As shown by
Molinari et al. (\cite{molinari}) and Fontani et al. (\cite{fontani}) the flux 
densities at these 
wavelengths usually arise from regions outside the core that are detected at
the longer wavelengths, and one needs at least 2 grey-body fits to cover the
entire SED. The SED-fit shown in Fig.~\ref{sed} was obtained by assuming 
a dust opacity $k_{\nu} = k_{250}(\nu[GHz] / 250)^{\beta}$, with 
$k_{250} = 1$~cm$^2$g$^{-1}$ (Ossenkopf \& Henning~\cite{ossenkopf}). 
We obtained a good fit with $\beta=1.8$, and a 
dust temperature $T_{\rm dust}=23$~K.
The resulting dust mass is 11~\Msol. 
For a gas-to-dust ratio of 100, the total mass of the core detected in the
(sub-)mm continuum is 1100~\Msol, consistent with what is found from the 
$^{13}$CO and C$^{18}$O-data (see Table~\ref{cloudlineparams}).
We should keep in mind, however, that the dust properties and the gas-to-dust 
ratio in the FOG may differ from those found near the Sun.

\section{The distance to WB89-789}

The presence of
an outflow, together with the H$_2$O maser emission detected by Wouterloot 
et al. (\cite{wbf}) demonstrates that star formation is still active in 
this object. If the kinematic distance of 11.9~kpc can be confirmed, this
would be the 
star-forming region furthest from the Galactic center yet found in the Galaxy.
The distance was derived from the $^{12}$CO velocity, using the Brand \& Blitz
(\cite{bb93}) rotation curve. 
WB89-789 is at a galactic longitude of 195\pad82; 
the determination of kinematic distances is
notoriously uncertain in directions near the (anti-) center, where 
$V_{\rm lsr}$ is not very sensitive to distance, and random motions become 
relatively influential in the distance determination. 
If we assume a cloud-cloud velocity dispersion of $\sim$5~\kms (see Brand \& 
Blitz~\cite{bb93}), then the range of galactocentric distance of WB89-789 is 
$17\lsim R \lsim 25$~kpc (with a corresponding range in heliocentric distance
$8.6\lsim d \lsim 17.1$~kpc).

\begin{figure}[tp]
\resizebox{8cm}{!}{\rotatebox{270}{
\includegraphics{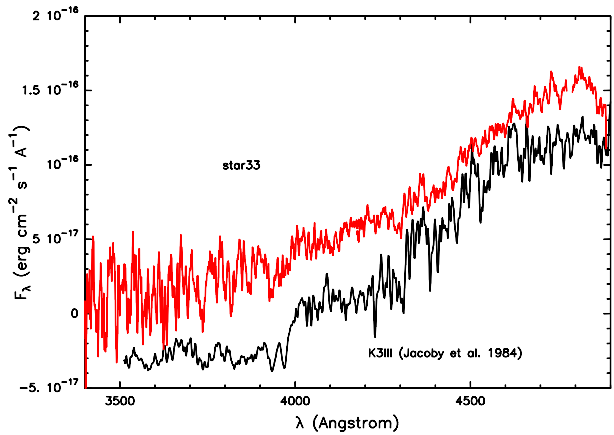}}}
\caption[]{
Spectrum of star~33 (see Fig.~\ref{NIRid1}) taken with DOLORES at the TNG. The
spectrum has been smoothed to a resolution of $\sim 5.4$\AA. Also shown is 
the spectrum of a K3III-star (from Jacoby et al.~\cite{jacoby}, arbitrarily 
shifted vertically to allow comparison). Because of the
poor wavelength calibration, the spectrum of star~33 has been shifted in 
$\lambda$ to get correspondence between the features in both stars. The amount 
of shifting necessary agrees with that which we determined from a comparison of
the lines in our flux calibrator (Hiltner~600) and Jacoby et al. 
(\cite{jacoby}) star
No.13, which has a similar spectral type.
}
\label{dolores}
\end{figure}

\smallskip
As shown in Fig.~\ref{colmag}a, star~33 (indicated in Fig.~\ref{NIRid1}), 
if it is a main sequence star, would be its earliest type member that does 
not have NIR excess, and would have a spectral type $\sim$B0, 
if the cluster is at a distance of 11.9~kpc. 
To check this, we have obtained a spectrum of this star 
(see Fig.~\ref{dolores}). 
From the way the continuum rises towards longer wavelengths it is
clear that star~33 is not an early-type star. 
We have compared the TNG-spectrum with those in the library of 
Jacoby et al. (\cite{jacoby}). To reduce the noise we smoothed the spectrum 
to a resolution of 5.4\AA, comparable to the 4.5\AA-resolution of the
Jacoby et al. (\cite{jacoby}) spectra. From a visual inspection we conclude 
that the spectrum
of star~33 agrees best with that of a K3~III-star (Jacoby et al. 
(\cite{jacoby}) 
spectrum No.~100; this spectrum is also shown in Fig.~\ref{dolores}). 
This implies that star~33 has a visual extinction 
A$_{\rm V} \sim 5.5$, and a geocentric distance of 10.7~kpc, i.e., 
$R \approx 19.0$~kpc. Identification as a K3~III-star is consistent with 
the location of star~33 in the color-magnitude diagram of Fig.~\ref{colmag}:
dereddening brings it first on the giant branch, at around spectral type K3.

\smallskip
This result is rather surprising: star~33 is embedded in nebulosity (cf. 
Fig.~\ref{NIRid1}a), 
making it a likely member of the cluster associated with the
IRAS source. 
On the other hand, a K3III-star would originate from a 9--10~\Msol\ ms-star 
(i.e., spectral type B2--B3), and has an age of about 20--25~Myr. 
It is possible that star formation has been going
on in this region for a much longer time, and that the IRAS source, the water
maser, the outflow, and the NIR-excess stars are just the latest episode. 
The earlier occurrence of star formation (giving rise to star~33) would have
removed much of the original molecular cloud, perhaps explaining why the 
cloud in which star formation is currently taking place is so small. 

If this scenario is correct, then it still has to be explained how star 
formation can be sustained over such a long time, especially in this part of
the Galaxy. As mentioned in the 
Introduction, if star formation is to be initiated by external triggers, then
it is difficult to see how this can work efficiently in the far-outer Galaxy,
where external triggers (SN explosions, spiral arms) are much more scarce or
weak. On the other hand, as already found by Wouterloot et al. (\cite{wbh}) and
Snell et al. (\cite{snell}), 
once star formation does occur in FOG-clouds, it appears to proceed as it
does locally. 
It should also be pointed out that there are other examples of older stars 
located in younger clusters. Kumar et al. (\cite{kumar}) found three late-type
giants in two young (1.5$-$1.9~Myr) open clusters, indicating that star 
formation has been going on there for about 25~Myr.
However, the spectral classification of star~33 is very tentative, and 
further speculation should wait until a better spectrum is available.

\section{Conclusions}

We have studied the star-forming region associated with 
IRAS06145+1455 (WB89-789) in detail.
Its location near the edge of the molecular disk (at a galactocentric distance 
of 20.2~kpc) makes this the 
galactic star forming region 
furthest from the Galactic center found up to now.
An embedded cluster containing about 60 stars was found in JHK-images.
The radius of the cluster is about 1.3~pc, within which the average stellar
surface density is 12~pc$^{-2}$; within a core radius of 0.4~pc the stellar
surface density is 33~pc$^{-2}$. The stars have a
foreground extinction of about 6~mags and an internal extinction of
up to 30~mags, though typically $\sim 7.5$~mags. 
At least 14 of the embedded stars show NIR-color excess in a J$-$H vs. H$-$K 
diagram, suggestive of the presence of disks around these stars; three of 
these are possibly of Class~I, while the locations of the others suggest they 
are Class~II objects.

An optical spectrum of what seemed like it could be the earliest type cluster 
member with normal extinction shows a tentative likeness to that of  
a K3~III star. The distance of this star would then be 
10.7~kpc (galactocentric distance 19.0~kpc), consistent with the
kinematical distance of the region. The presence of this star in the
cluster suggests that star formation has proceeded for the past several 
10$^7$ years.
The absence of radio continuum emission towards this source, at the sensitivity
of our observations, indicates that no star of type B0.5~V or earlier is 
present.

JCMT $^{13}$CO and C$^{18}$O and IRAM CS observations show that the cluster is
embedded in a $\sim 500-1500$~\Msol\ ($M_{\rm vir}$) core. 
A JCMT $^{12}$CO map shows 
that the core is contained in a $\sim 5000$~\Msol\  cloud, consisting of two 
clumps. The clump with the embedded cluster shows the
existence of moderate outflow emission. (Sub-)mm continuum maps (JCMT 
and SEST) at 0.45, 0.85, and 1.2~mm show the presence of a dust core 
centered at the cluster. With standard assumptions the dust mass is 10~\Msol;
assuming the canonical gas/dust ratio of about 100, the total core mass is 
$\sim 1000$~\Msol, which is consistent with what is found from $^{13}$CO and 
C$^{18}$O data.
The question now is which mechanism started the star formation process in
this cloud, and how it can be sustained over such a long time (i.e., at least 
$20-25$~Myr, the age of a K3~III-star).

\acknowledgements
This work is partly based on observations collected at
the European Southern Observatory, Chile.
The James Clerk Maxwell Telescope is operated by the Joint Astronomy Centre 
on behalf of the Particle Physics and Astronomy Research
Council of the United Kingdom, The Netherlands Organisation for Scientific 
Research, and the National Research Council of Canada. Observations were 
carried out under projects nls0203, m02bn13, m03bn22, and m05bn27b.
\noindent
This paper is partially based on observations made with 
the Italian Telescopio Nazionale Galileo (TNG). The TNG is operated on the 
island of
La Palma by the Fundaci\'on Galileo Galilei of the INAF (Istituto Nazionale di
Astrofisica) at the Spanish Observatorio del Roque de los Muchachos of the
Instituto de Astrofisica de Canarias.
\noindent
This research has made use of NASA's Astrophysics Data System Bibliographic 
Services (ADS) and of the SIMBAD database, operated at CDS, Strasbourg, 
France. Equatorial coordinates of the NIR-objects were determined using the 
GAIA software. 
\noindent
We thank Robert Zylka for his help with the SIMBA data reduction procedures
and scripts, and Giovanna Stirpe for help with the reduction of the 
spectroscopic data. The spectra were taken at the TNG in Service Mode by 
Gloria Andreuzzi, to whom we are grateful. We thank Francesco Palla 
for helpful discussions on Sect.~4. We are 
indebted to Loris Magnani and Riccardo Cesaroni for commenting on an earlier 
version of this paper.

\clearpage

\begin{table*}
\caption{NIR photometry data}
\label{photomdata}
\begin{flushleft}
\begin{tabular}{rrrrrrrrrrr}
\hline\noalign{\smallskip}
\multicolumn{1}{c}{nr} & \multicolumn{1}{c}{$\alpha (2000)$}  
 & \multicolumn{1}{c}{$\delta (2000)$} 
 & \multicolumn{1}{c}{K} 
 & \multicolumn{1}{c}{$\sigma_{\rm K}$}
 & \multicolumn{1}{c}{H} & \multicolumn{1}{c}{$\sigma_{\rm H}$}
 & \multicolumn{1}{c}{J} & \multicolumn{1}{c}{$\sigma_{\rm J}$} 
 & \multicolumn{1}{c}{H--K} & \multicolumn{1}{c}{J--H} \\
\multicolumn{1}{c}{} & \multicolumn{1}{c}{$h$  $m$  $s$} & 
\multicolumn{1}{c}{$\circ$  \arcmin\  \arcsec} & \multicolumn{8}{c}{} \\
\hline\noalign{\smallskip}
   3& 06:17:25.79& +14:55:00.1 & 16.44 & 0.09 &  17.57 & 0.16 &  18.70 & 0.23 & 1.125 & 1.128\\
   4& 06:17:25.73& +14:54:50.9 & 16.57 & 0.10 &  17.12 & 0.09 &  18.21 & 0.16 & 0.556 & 1.083\\
   5& 06:17:25.57& +14:55:05.4 & 13.06 & 0.04 &  13.98 & 0.05 &  15.46 & 0.02 & 0.912 & 1.479\\
   7& 06:17:25.42& +14:54:28.7 & 16.22 & 0.14 &  16.74 & 0.12 &  17.12 & 0.09 & 0.518 & 0.383\\
   8& 06:17:25.42& +14:54:23.8 & 15.56 & 0.05 &  16.21 & 0.08 &  17.24 & 0.08 & 0.644 & 1.035\\
   9& 06:17:25.37& +14:54:59.7 & 17.65 & 0.20 &  18.60 & 0.29 &  19.20 & 0.39 & 0.948 & 0.597\\
  10& 06:17:25.33& +14:54:30.7 & 15.58 & 0.06 &  16.05 & 0.04 &  17.25 & 0.10 & 0.472 & 1.200\\
  11& 06:17:25.28& +14:54:46.9 & 16.80 & 0.22 &  17.88 & 0.23 &  19.03 & 0.36 & 1.072 & 1.154\\
  12& 06:17:25.24& +14:54:49.3 & 14.27 & 0.07 &  15.30 & 0.06 &  16.60 & 0.05 & 1.032 & 1.307\\
  13& 06:17:25.24& +14:54:33.0 & 13.25 & 0.04 &  13.50 & 0.04 &  13.78 & 0.01 & 0.253 & 0.277\\
  14& 06:17:25.12& +14:54:39.2 & 15.16 & 0.05 &  15.92 & 0.06 &  17.32 & 0.08 & 0.753 & 1.406\\
  15& 06:17:25.10& +14:54:56.5 & 13.39 & 0.04 &  15.04 & 0.04 &  16.74 & 0.05 & 1.647 & 1.694\\
  16& 06:17:25.12& +14:54:52.2 & 16.70 & 0.14 &  17.75 & 0.14 &  19.17 & 0.38 & 1.045 & 1.429\\
  17& 06:17:25.07& +14:54:21.1 & 13.59 & 0.06 &  13.83 & 0.19 &  14.31 & 0.08 & 0.239 & 0.488\\
  20& 06:17:24.95& +14:54:33.9 & 17.17 & 0.20 &  17.69 & 0.15 &  19.15 & 0.47 & 0.525 & 1.457\\
  21& 06:17:24.95& +14:54:26.6 & 17.30 & 0.21 &  18.16 & 0.19 &  19.12 & 0.37 & 0.859 & 0.967\\
  23& 06:17:24.84& +14:54:36.9 & 16.62 & 0.12 &  17.34 & 0.09 &  18.69 & 0.24 & 0.715 & 1.346\\
  24& 06:17:24.81& +14:55:18.7 & 14.66 & 0.04 &  15.13 & 0.03 &  16.28 & 0.06 & 0.469 & 1.153\\
  25& 06:17:24.78& +14:54:55.9 & 15.81 & 0.07 &  16.63 & 0.06 &  17.97 & 0.12 & 0.825 & 1.334\\
  26& 06:17:24.78& +14:54:53.8 & 14.58 & 0.03 &  15.44 & 0.04 &  16.91 & 0.06 & 0.855 & 1.468\\
  27& 06:17:24.75& +14:54:59.1 & 15.20 & 0.07 &  15.93 & 0.07 &  16.93 & 0.09 & 0.732 & 1.002\\
  28& 06:17:24.78& +14:54:32.5 & 17.09 & 0.15 &  17.48 & 0.10 &  18.38 & 0.18 & 0.392 & 0.898\\
  29& 06:17:24.50& +14:54:22.7 & 17.02 & 0.19 &  17.73 & 0.18 &  18.68 & 0.25 & 0.706 & 0.949\\
  30& 06:17:24.45& +14:55:10.8 & 16.74 & 0.14 &  17.09 & 0.09 &  17.95 & 0.16 & 0.348 & 0.858\\
  31& 06:17:24.43& +14:54:48.3 & 14.87 & 0.05 &  15.64 & 0.05 &  17.02 & 0.06 & 0.770 & 1.382\\
  33& 06:17:24.33& +14:54:38.0 & 13.07 & 0.04 &  13.55 & 0.12 &  14.54 & 0.01 & 0.476 & 0.987\\
  34& 06:17:24.36& +14:54:28.5 & 15.83 & 0.10 &  16.31 & 0.08 &  17.34 & 0.10 & 0.477 & 1.026\\
  35& 06:17:24.30& +14:54:48.2 & 14.57 & 0.05 &  15.69 & 0.06 &  17.28 & 0.09 & 1.120 & 1.583\\
  36& 06:17:24.28& +14:54:53.0 & 16.63 & 0.14 &  17.40 & 0.11 &  19.17 & 0.34 & 0.769 & 1.769\\
  37& 06:17:24.24& +14:54:33.1 & 13.65 & 0.05 &  14.44 & 0.04 &  15.74 & 0.04 & 0.786 & 1.301\\
  40& 06:17:23.96& +14:54:32.5 & 13.14 & 0.13 &  13.56 & 0.05 &  14.12 & 0.01 & 0.422 & 0.560\\
  42& 06:17:23.85& +14:55:12.9 & 16.74 & 0.11 &  17.56 & 0.12 &  18.76 & 0.24 & 0.819 & 1.207\\
  44& 06:17:23.76& +14:55:17.8 & 13.80 & 0.08 &  14.45 & 0.09 &  15.58 & 0.07 & 0.654 & 1.126\\
  45& 06:17:23.78& +14:54:34.0 & 13.94 & 0.18 &  14.66 & 0.20 &  15.99 & 0.05 & 0.713 & 1.334\\
\noalign{\smallskip}
\hline
\end{tabular}
\end{flushleft}
\end{table*}

\addtocounter{table}{-1}

\begin{table*}
\caption{{\it continued}}
\begin{flushleft}
\begin{tabular}{rrrrrrrrrrr}
\hline\noalign{\smallskip}
\multicolumn{1}{c}{nr} & \multicolumn{1}{c}{$\alpha (2000)$}  
 & \multicolumn{1}{c}{$\delta (2000)$}
 & \multicolumn{1}{c}{K} 
 & \multicolumn{1}{c}{$\sigma_{\rm K}$}
 & \multicolumn{1}{c}{H} & \multicolumn{1}{c}{$\sigma_{\rm H}$}
 & \multicolumn{1}{c}{J} & \multicolumn{1}{c}{$\sigma_{\rm J}$} 
 & \multicolumn{1}{c}{H--K} & \multicolumn{1}{c}{J--H} \\
\multicolumn{1}{c}{} & \multicolumn{1}{c}{$h$ $m$ $s$} & 
\multicolumn{1}{c}{$\circ$ \arcmin\  \arcsec} & \multicolumn{8}{c}{} \\ 
\hline\noalign{\smallskip}
  46& 06:17:23.79& +14:54:27.9 &16.25 & 0.10 &  16.74 & 0.08 &  17.78 & 0.12 & 0.487 & 1.037\\
  47& 06:17:23.64& +14:54:36.3 &15.74 & 0.09 &  16.26 & 0.07 &  17.45 & 0.12 & 0.528 & 1.190\\
  48& 06:17:23.56& +14:55:15.8 &17.32 & 0.16 &  17.70 & 0.12 &  18.68 & 0.23 & 0.378 & 0.978\\
  49& 06:17:23.56& +14:54:52.8 &17.06 & 0.16 &  18.32 & 0.21 &  19.60 & 0.52 & 1.269 & 1.280\\
  50& 06:17:23.54& +14:54:50.1 &16.27 & 0.08 &  16.96 & 0.09 &  18.12 & 0.15 & 0.691 & 1.157\\
  51& 06:17:23.53& +14:54:39.6 &16.15 & 0.08 &  16.50 & 0.09 &  17.71 & 0.11 & 0.348 & 1.212\\
  53& 06:17:23.46& +14:55:06.6 &16.96 & 0.13 &  17.39 & 0.12 &  18.83 & 0.29 & 0.433 & 1.433\\
  54& 06:17:23.42& +14:54:40.0 &16.19 & 0.07 &  18.29 & 0.26 &  19.34 & 0.43 & 2.097 & 1.053\\
  56& 06:17:23.30& +14:55:14.9 &15.42 & 0.05 &  15.60 & 0.11 &  15.95 & 0.14 & 0.173 & 0.350\\
  57& 06:17:23.32& +14:54:23.0 &16.14 & 0.09 &  16.75 & 0.06 &  17.80 & 0.11 & 0.605 & 1.057\\
  60& 06:17:23.23& +14:54:28.2 &15.52 & 0.08 &  16.25 & 0.07 &  17.57 & 0.13 & 0.730 & 1.318\\
  62& 06:17:23.14& +14:54:36.9 &16.25 & 0.08 &  16.79 & 0.06 &  18.02 & 0.13 & 0.544 & 1.224\\
  63& 06:17:23.08& +14:54:31.8 &17.54 & 0.24 &  18.86 & 0.42 &  19.63 & 0.54 & 1.325 & 0.762\\
  64& 06:17:23.02& +14:54:58.0 &13.68 & 0.07 &  13.97 & 0.07 &  14.58 & 0.22 & 0.288 & 0.614\\
  65& 06:17:23.05& +14:54:22.0 &16.52 & 0.11 &  17.01 & 0.06 &  18.14 & 0.14 & 0.487 & 1.130\\
  66& 06:17:22.82& +14:55:13.1 &15.74 & 0.06 &  16.33 & 0.05 &  17.33 & 0.07 & 0.581 & 1.006\\
  68& 06:17:22.75& +14:54:59.5 &14.12 & 0.06 &  14.59 & 0.12 &  15.70 & 0.08 & 0.472 & 1.110\\
  69& 06:17:22.70& +14:54:41.6 &13.09 & 0.06 &  13.63 & 0.04 &  14.67 & 0.03 & 0.542 & 1.045\\
  70& 06:17:22.64& +14:54:56.4 &14.99 & 0.06 &  15.19 & 0.03 &  15.88 & 0.13 & 0.201 & 0.693\\
  71& 06:17:22.64& +14:54:33.8 &15.72 & 0.05 &  15.89 & 0.09 &  16.40 & 0.14 & 0.166 & 0.512\\
  72& 06:17:22.45& +14:54:55.0 &15.79 & 0.07 &  16.23 & 0.05 &  17.28 & 0.08 & 0.435 & 1.052\\
  76& 06:17:25.10& +14:54:22.4 &15.89 & 0.08 &  16.21 & 0.09 &  16.78 & 0.07 & 0.327 & 0.562\\
  80& 06:17:24.25& +14:54:44.2 &16.58 & 0.23 &  17.18 & 0.13 &  18.10 & 0.15 & 0.607 & 0.915\\
  81& 06:17:24.28& +14:54:26.6 &16.76 & 0.25 &  17.79 & 0.25 &  18.38 & 0.25 & 1.027 & 0.593\\
  83& 06:17:23.95& +14:54:41.8 &14.74 & 0.28 &  17.02 & 0.27 &  18.11 & 0.19 & 2.287 & 1.091\\
  86& 06:17:23.68& +14:55:18.3 &17.03 & 0.22 &  17.67 & 0.21 &  18.94 & 0.40 & 0.639 & 1.275\\
  87& 06:17:23.37& +14:54:42.2 &17.75 & 0.32 &  17.83 & 0.16 &  19.22 & 0.40 & 0.085 & 1.384\\
  88& 06:17:23.13& +14:54:26.6 &16.24 & 0.19 &  16.49 & 0.08 &  17.13 & 0.09 & 0.254 & 0.638\\
  89& 06:17:23.09& +14:54:57.2 &16.54 & 0.18 &  17.28 & 0.16 &  18.81 & 0.47 & 0.739 & 1.535\\
  90& 06:17:22.79& +14:54:41.7 &15.11 & 0.07 &  15.22 & 0.05 &  15.69 & 0.04 & 0.105 & 0.471\\
  92& 06:17:24.82& +14:54:51.7 &15.17 & 0.09 &  16.39 & 0.09 &  18.30 & 0.20 & 1.220 & 1.913\\
  93& 06:17:24.76& +14:54:51.4 &15.61 & 0.13 &  15.72 & 0.08 &  16.35 & 0.06 & 0.107 & 0.634\\
  98& 06:17:23.99& +14:54:35.1 &16.57 & 0.18 &  17.12 & 0.23 &  17.84 & 0.21 & 0.541 & 0.723\\
  99& 06:17:23.73& +14:54:34.7 &16.59 & 0.25 &  17.46 & 0.32 &  17.43 & 0.18 & 0.869 &-0.033\\
\noalign{\smallskip}
\hline
\end{tabular}
\end{flushleft}
\end{table*}

\end{document}